\documentclass[10pt,aps,prl,showpacs,twocolumn,superscriptaddress,floats,floatfix]{revtex4}
\usepackage{amssymb,amsmath}
\usepackage{graphics}
\usepackage{color}
\usepackage{epstopdf}
\usepackage{epsfig}
\usepackage{rotating}
\usepackage{float}
\usepackage{bm}
\usepackage{soul}
\usepackage{hyperref}
\usepackage{breakurl}
\usepackage[stable,perpage]{footmisc}
\usepackage{xr}
\newcommand{\rem}[1]{}

\begin{document}
\title{Binary-Fluid Turbulence: Signatures of Multifractal Droplet Dynamics and Dissipation Reduction}
\author{Nairita Pal\footnote{nairitap2009@gmail.com}}
\affiliation{Centre for Condensed Matter Theory, Department of Physics, Indian Institute 
of Science, Bangalore 560012, India.}
\author{Prasad Perlekar\footnote{perlekar@tifrh.res.in}}
\affiliation{TIFR Centre for Interdisciplinary Sciences, 21 Brundavan Colony, Narsingi, Hyderabad 500075, India}
\author{Anupam Gupta\footnote{anupam1509@gmail.com}}
\affiliation{Department of Physics and INFN, University of ``Tor Vergata'', Via della Ricerca Scientifica 1, 00133 Rome, Italy.}
\author{Rahul Pandit\footnote{rahul@physics.iisc.ernet.in; 
also at Jawaharlal Nehru Centre For
Advanced Scientific Research, Jakkur, Bangalore, India.}
}
\affiliation{Centre for Condensed Matter Theory, Department of Physics, Indian Institute 
of Science, Bangalore 560012, India.}

\date{\today} 

\begin{abstract}
We study the challenging problem 
of the advection of an active, deformable,
finite-size droplet by a turbulent flow 
via 
a simulation of the coupled Cahn-Hilliard-Navier-Stokes (CHNS) equations.
In these equations,
the droplet has
a natural two-way coupling to the background fluid.
We show that the probability distribution function of the
droplet center of mass acceleration components
exhibit
wide, non-Gaussian tails, which are
consistent with the predictions based on pressure spectra.
We also show that the droplet deformation
displays multifractal dynamics.
Our study reveals that the presence of the droplet enhances the
energy spectrum $E(k)$, when the wavenumber $k$ is large; this enhancement leads
to dissipation reduction. 

\end{abstract}

\maketitle
\section{I. INTRODUCTION}
The advection of droplets, bubbles, or particles by a fluid plays a central
role in many natural and industrial settings~\cite{jeremie2006}, which include
clouds~\cite{grabowski2013,shaw2003}, fuel injection~\cite{fuel},
microfluidics~\cite{baroud2010}, inkjet printing~\cite{inkjet}, and the
reduction of drag by bubbles~\cite{lohse2003}. 
These studies require an accurate modeling
of the motion of particles or droplets inside a turbulent fluid.
The advection of
finite-sized particles or droplets is especially challenging because
they cannot be 
modelled as Lagrangian tracers~\cite{bifarale2005}, or even 
like heavy-particles, which do not affect the motion of the carrier phase~\cite{jeremie2006}.

Finite-size, deformable droplets affect the background fluid considerably,
even as they are transported and deformed by the flow.
This makes a systematic characterization
of the statistical properties of turbulence difficult, because 
boundary conditions have to be implemented
on the surface of the droplet, which changes as a function of time.
The Cahn-Hilliard-Navier-Stokes (CHNS) equations that we use allow us to treat droplets
elegantly via gradients in an order-parameter field $\phi$; therefore, we do not
have to enforce complicated boundary conditions at the moving
boundary between the droplet and the background fluid; and,
we can follow the deformation of the droplet boundary in far greater 
detail than has been possible so far. Our ability to track this boundary, along
with our efficient computer code on a GPU cluster has enabled us to show, among other things,
that fluctuations of the droplet
boundary are multifractal; this has not been investigated
hitherto.

The simplest droplet-advection
problem arises in a binary-fluid mixture, in which a droplet of the minority
phase moves in the majority-phase background that is turbulent. We study this problem
in two spatial dimensions (2D)
by using the coupled CHNS equations, 
which have been used extensively in studies of  critical
phenomena, phase
transitions~\cite{chaikan2000,halperin77,lifshitz59,domb83,bray94},
nucleation~\cite{nucleation}, spinodal decomposition~\cite{onuki2002,
badalassi2003,perlekar2014,cahn61,bofetta2005}, and the late stages of phase
separation~\cite{bofetta2009}. 
We use the CHNS
approach to
 carry out a
detailed study of droplet dynamics in a turbulent flow and characterize the
turbulence-induced deformation of a droplet and its acceleration statistics.
We then elucidate the modification of fluid turbulence by the fluctuations of
this droplet. 
Our study
uses an extensive direct numerical simulation (DNS) of
the CHNS equations in 2D, where we
use parameters such that we have one droplet in our
simulation domain.
We track such a finite-sized
droplet (for similar studies of Lagrangian or inertial particles see
Ref.~\cite{prasad2011}) and obtain the statistics of the deformation of the
droplet and its velocity and acceleration statistics as a function of the
surface tension and size.

2D fluid turbulence, which is of central importance in many
flows, is fundamentally different from its three-dimensional (3D)
counterpart~\cite{fjortoff,Kraich,Leith,Batchelor,leisure}. The
fluid-energy spectrum $E(k)$ in 2D turbulence shows (a) a {\it forward cascade} of
enstrophy (or the mean-square vorticity), from the forcing wave number $k_{f}$
to wave numbers $k > k_{f}$ and (b) an {\it {inverse cascade}} of energy to 
$k < k_{f}$.  We use parameters that lead to an $E(k)$
that is dominated by a forward-cascade regime.  Our study leads to new
insights and remarkable results: 
we show that the turbulence-induced
fluctuations in the dimensionless deformation of the droplet are
intermittent.
We
characterize this intermittency of the droplet fluctuations by obtaining the
probability distribution function (PDF) $P_{\Gamma}(\Gamma)$ and the
multifractal spectrum $f_{\Gamma}(\alpha)$ of the time series $\Gamma(t)$. We
show that the PDF of the components of the acceleration of the center of mass
are similar to those for finite-size particles in turbulent
flows~\cite{jeremie} and are consistent with predictions based on pressure
spectra~\cite{hill,qureshi2007}.  We also find that the large-$k$
tail of $E(k)$ is enhanced by the droplet
fluctuations; this leads to dissipation reduction, in much the same
way as in turbulent fluids with polymer additives~\cite{kalelkar05, perlekar06,
gupta15}.  The spectrum $E(k)$ also displays oscillations whose period is
related inversely to the mean diameter of the droplet. We show that such
oscillations appear prominently in the order-parameter spectrum $S(k)$, which is the
Fourier transform of the spatial correlation function of $\phi$, the
Cahn-Hilliard scalar field that distinguishes between the two binary-fluid phases.

The remainder of the paper is organized as follows. Section II introduces
the CHNS equations and the numerical methods we use to solve them.
We present the results of our DNS in Section III, which
comprises subsections on 
(a) droplet-deformation statistics, (b)
droplet-acceleration statistics, (c) energy-dissipation
time series
and energy and order-parameter spectra.
Section IV contains a discussion of our results
and conclusions. An Appendix contains some details
of our calculations. 
\section{II. MODEL AND NUMERICAL METHODS}

Two-way coupling, between the droplet and the background turbulent fluid,
appears naturally in the CHNS
equations~\cite{celani2009,scarbolo2013,shen2004,scarbolo2011}. In
2D, the Navier-Stokes equations can be written
in the following stream-function vorticity formulation
~\cite{bofetta2009}: 
\begin{eqnarray}
 \left(\partial_t + {\bm u}\cdot \nabla\right) \omega &=& \nu \nabla^2 \omega - \alpha \omega-\nabla \times (\phi \nabla \mu) + F_\omega ;
 \label{ns}\\ 
 \left(\partial_t + {\bm u}\cdot \nabla\right){\bf \phi} & = & \gamma \nabla^2 {\mu}~{\rm{and}}~ \nabla \cdot {\bm u} =0.
 \label{ch}
 \end{eqnarray}
Here ${\bm u}\equiv(u_x,u_y)$ is the fluid velocity, $\omega$
is the vorticity, $\mu$ is the chemical potential; $\omega$
and $\mu$ are connected to
${\bm u}$ and $\phi$ in the following way:
\begin{eqnarray}
\omega&=&(\nabla \times {\bm u}) {\hat{\bm e}}_z,\\ 
\mu({\bm x},t) & = & \delta {\mathcal F}[\phi]/\delta \phi,\\
{\mathcal F}[\phi] & = &\Lambda \int
[(\phi^2-1)^2/(4\xi^2) \nonumber \\
&&+ |\nabla \phi|^2/2] d{\bm x}
\phi({\bm x},t),
\end{eqnarray}
 where ${\mathcal F}[\phi]$ is the free energy.
In Eqs.~(\ref{ns})-(\ref{ch})
$\Lambda$ is the energy density with which the two phases mix in the
interfacial regime~\cite{celani2009}, $\xi$ sets the scale of the diffuse interface
width, $\nu$ is the kinematic viscosity, $\gamma$ is
the mobility~\cite{shen2004} of the binary-fluid
mixture, $F_{\omega}=F_0\cos(k_f y)$ is a Kolmogorov-type
forcing~\cite{kolmogorov} with amplitude $F_0$ and forcing wave number $k_f$,
and $\alpha$ is the air-drag induced friction. For simplicity, we concentrate
on mixtures in which  $\gamma$ is independent of $\phi$ and both components
have the same density and viscosity. In our model,
$\sigma=\frac{2\sqrt{2}\Lambda}{3\xi}$
is the surface tension. The Grashof number
$Gr=\frac{L^{4}F_{0}}{\nu^{2}}$ is a convenient dimensionless ratio of the forcing
and viscous terms.  We keep the diffusivity $D=\frac{\gamma\beta}{\xi^{2}}$ of
the system constant. The forcing-scale Weber number $We\equiv \rho
L_f^{3}F_0/\sigma$, where $L_f=2\pi/k_f$, is a natural dimensionless measure of
the inverse of the surface tension.

The minority and majority phases in our model
 is described by
an order-parameter field $\phi({\bm x},t)$ 
at the point ${\bm x}$ and time $t$ with $\phi({\bm
x},t)>0$ in the background (majority) phase and $\phi({\bm x},t)<0$ in the
droplet (minority) phase (see Fig.~\ref{def}(a)).
At time $t=0$ we begin with the order-parameter profile~\cite{celani2009,scarbolo2011}
\begin{equation}
\phi(x,y)=\tanh\left[\frac{1}{\sqrt{2}\xi}\left(\sqrt{(x-x_c)^{2}+(y-y_c)^{2}}-d_0/2\right)\right] ,
\label{tanh}
\end{equation}
which ensures that
the droplet is circular at $t=0$, with its
center at $(x_c, y_c)$, diameter $d_0$, and  
 has a diffuse interface, because
$\phi$ change {\it{continuously}} in the interface.
The interface width $\xi$ is measured by the
dimensionless Cahn number $Ch=\xi/L$.

  Our direct numerical simulations (DNSs)
of Eqs.~\eqref{ns} and \eqref{ch} use a pseudospectral method and periodic
boundary conditions; $L(=2\pi)$ is
the linear size of our square simulation domain
which has $N^2$
collocation points. We
have a cubic nonlinearity in the chemical
potential $\mu$ (Eq.~\ref{ch}), so
we use $N/2$-dealiasing~\cite{spectral}.  For time integration
we use the exponential Adams-Bashforth method ETD2~\cite{cox2002}. We use
computers with Graphics Processing Units (e.g., the NVIDIA K80), which we
program in CUDA~\cite{cuda}; our efficient code allows us to explore the CHNS
parameter space and carry out very long simulations that are essential for our
studies. In the following paragraph we introduce the quantities
that we calculate from the fields $\omega({\bm x},t)$ and $\phi({\bm x},t)$,
which we
obtain from our DNSs of Eqs.~\eqref{ns} and~\eqref{ch}.

From the field $\phi({\bm x},t)$ we calculate the 
droplet deformation parameter which we define as~\cite{perlekar12},
\begin{equation}
\Gamma(t)=\frac{{\mathcal {S}}(t)}{{\mathcal {S}}_{0}(t)}-1,
\label{peri}
\end{equation}
where
${\mathcal {S}}(t)$ is the perimeter of the droplet (the $\phi=0$ contour) at time $t$,
${\mathcal {S}}_0(t)$ is the perimeter of an undeformed droplet of equal area at $t$.
From the field $\omega({\bm x},t)$ we calculate the
total kinetic energy of the fluid $E(t)$, and the fluid-energy dissipation
rate $\varepsilon(t)$, which are
\begin{eqnarray} 
E(t) & = & \langle |{\bm u}({\bm x},t)|^{2} \rangle_{{\bm x}},\\ 
\varepsilon(t) & = & \langle \nu |\omega({\bm x},t)|^{2}
\rangle_{{\bm x}},
\end{eqnarray} 
where $\langle \rangle_{{\bm x}}$ denotes the average over
space. From
$E(t)$ and $\varepsilon(t)$ we
calculate the root-mean-square fluid velocity,
$u_{rms}=\sqrt{\langle 2E(t) \rangle_t}$,
where $\langle \rangle_t$ denotes the average over the statistically steady, but turbulent
state with small fluctuations
about the mean value, i.e, the fluid is in
the {\it{statistically stationary}} state. From these, we  
calculate the Taylor-microscale Reynolds 
number $Re_{\lambda}(t)=\sqrt{2}E(t)/\sqrt{\nu\varepsilon(t)}$, 
and the mean $\langle Re_{\lambda}\rangle_t$, which
characterizes the intensity of turbulence and
the box-size eddy-turnover time $\tau_{eddy}=L/u_{rms}$; we express time
in units of $\tau_{eddy}$.
We calculate the
energy spectra $E(k)$ and order-parameter (or phase-field) spectra $S(k)$
as follows:
\begin{eqnarray}
E(k) & \equiv & \sum \limits_{k-\frac{1}{2} \leq k' \leq k+\frac{1}{2}} \langle
|\hat{\bm u}({\mathbf k}',t)|^{2} \rangle_{t},\\
S(k) & \equiv & \sum \limits_{k-\frac{1}{2}
\leq k' \leq k+\frac{1}{2}} \langle |\hat{\phi}({\mathbf k}',t)|^{2}
\rangle_{t},
\end{eqnarray}
where $\hat{\bm u}({\mathbf k}',t)$ and $\hat{\phi}({\mathbf k}',t)$
are, respectively, the spatial Fourier transforms of ${\bm u}({\bm x},t)$ and
$\phi({\bm x},t)$. 
 We have carried out
several DNSs ({\tt{R1}}-{\tt{R28}}) that
are given in Table I.
\begin{table*}
\resizebox{0.6\linewidth}{!}
{
\begin{tabular}{|l|l|l|l|l|l|l|l|l|l|l|l|}
\hline
&  $Gr$& $d_0/L$ & $We$ & $\langle d_p\rangle_t/L$& $\langle\lambda\rangle_t/L$& $\langle\eta\rangle_t/L$&$\langle E\rangle_t$ & $\langle\varepsilon\rangle_t$ &
$\langle Re_{\lambda}\rangle_t$ \\
\hline
\hline
{\tt R1} &  $3 \times 10^{7}$ &&&&&&$1.9$&$0.017$&$216$\\
\hline
{\tt R2} & $3 \times 10^{7}$ & $0.332$& $1.38$&  $0.324$& $0.08$&$0.007$ &$1.17$&$5.4$&$112$ \\
\hline
{\tt R3} & $3 \times 10^{7}$ &$0.312$& $1.38$&  $0.3$& $0.08$&$0.007$ &$1.24$&$5.1$&$120$ \\
\hline
{\tt R4} & $3 \times 10^{7}$ &$0.293$& $1.38$& $0.283$ & $0.09$&$0.007$&$1.3$&$4.9$&$127$ \\
\hline
{\tt R5} &  $3 \times 10^{7}$&$0.273$& $1.38$& $0.263$ &$0.09$&$0.007$&$1.36$&$0.023$&$137.5$ \\
\hline
{\tt R6} & $3 \times 10^{7}$ &$0.25$& $1.38$& $0.245$ &$0.09$&$0.007$&$1.4$&$4.4$&$146.5$ \\
\hline
{\tt R7}  & $3 \times 10^{7}$&$0.24$ & $5.34$&$0.2$ & $0.1$& $0.007$ &$1.4$&$4.63$&$140$ \\
{\tt R8}  &  $3 \times 10^{7}$&$0.24$& $2.3$& $0.22$&  $0.11$& $0.007$ &$1.44$&$4.35$&$151$\\
{\tt R9}  & $3 \times 10^{7}$&$0.24$& $1.97$& $0.22$&  $0.11$& $0.007$ &$1.45$&$4.2$&$153.4$\\
{\tt R10} &  $3 \times 10^{7}$&$0.24$& $1.84$& $0.22$&  $0.11$& $0.007$ &$1.48$&$4.25$&$154.7$\\
{\tt R11}  & $3 \times 10^{7}$ &$0.24$& $1.53$& $0.22$&  $0.11$& $0.007$ &$1.48$&$4.45$&$157.4$\\
{\tt R12} & $3 \times 10^{7}$&$0.24$& $1.38$& $0.22$&$0.12$& $0.007$ &$1.47$&$4.21$&$157$\\
{\tt R13}  & $3 \times 10^{7}$ &$0.24$& $0.534$& $0.22$&$0.12$&$0.007$&$1.5$&$4.19$&$160$\\
{\tt R14} &  $3 \times 10^{7}$&$0.24$& $0.138$& $0.22$& $0.12$&$0.007$ &$1.5$&$4.22$&$162$\\
\hline
{\tt R15} & $3 \times 10^{7}$ &$0.215$& $1.38$&$0.21$&$0.13$&$0.007$&$1.57$&$4.15$&$168$\\
\hline
{\tt R16} & $3 \times 10^{7}$&$0.2$& $1.38$&$0.177$&$0.13$&$0.007$&$1.62$&$3.96$&$177$\\
\hline
{\tt R17} &  $3 \times 10^{7}$&$0.174$& $1.38$&$0.153$&$0.14$&$0.007$&$1.7$&$3.8$&$188$\\
\hline
{\tt R18} & $3 \times 10^{7}$ &$0.14$&$5.34$ & $0.097$&$0.15$&$0.007$&$1.8$&$3.83$&$200$\\
{\tt R19}  &  $3 \times 10^{7}$&$0.14$& $2.3$&$0.125$&$0.15$&$0.007$&$1.75$&$3.83$&$195$\\
{\tt R20} &  $3 \times 10^{7}$&$0.14$& $1.38$&$0.126$ &$0.15$&$0.007$&$1.75$&$3.7$&$193$\\
\hline
{\tt R21} & $3 \times 10^{7}$ &$0.134$& $0.52$& $0.09$& $0.153$&$0.007$&$1.84$&$3.78$&$205$\\
\hline
{\tt R22} & $1.5 \times 10^{8}$ &&&& $0.12$ &$0.005$&$12.5$&$23.8$&$561.7$\\
\hline
{\tt R23} & $1.5 \times 10^{8}$ &$0.24$& $0.138$&$0.22$ & $0.094$ &$0.005$&$9.08$&$27.1$&$381.4$\\
\hline
{\tt R24} & $1.5 \times 10^{8}$ &$0.215$& $0.138$&$0.2$ & $0.1$&$0.005$&$9.5$&$25.4$&$411$\\
\hline
{\tt R25} & $1.5 \times 10^{8}$ &$0.2$& $0.138$&$0.176$ & $0.104$&$0.005$&$10.2$&$25.1$&$444$\\
\hline
{\tt R26} & $1.5 \times 10^{8}$ &$0.174$& $0.138$&$0.1525$ &$0.108$ &$0.005$&$10.7$&$23.9$&$477.8$\\
\hline
{\tt R27}  & $1.5 \times 10^{8}$ &$0.14$& $0.138$&$0.125$ &$0.112$ &$0.005$&$11.67$&$24.3$&$516.8$\\
\hline
{\tt R28} & $1.5 \times 10^{8}$ &$0.134$& $0.138$& $0.083$ &$0.116$ &$0.005$&$12.2$&$23.8$&$545.1$\\
\hline

\end{tabular}}
\caption{\label{table1} \small The parameters $Gr$, $d_0$, $We$, $\langle
d_p\rangle_t/L$, $\langle\lambda\rangle_t/L$, $\langle\eta\rangle_t/L$, $\langle E\rangle_t$, $\langle \varepsilon \rangle_t$,
and $\langle Re_{\lambda} \rangle_t$ for our DNS runs {\tt R1-R28}.  The number
of collocation points is kept fixed at $N^2=1024^2$ in each direction. The friction
coefficient $\alpha = 0.001$, the forcing wave number is fixed at $k_{f}=3$,
$\nu=4.67\times10^{-3}$ is the kinematic viscosity, the diffusivity
$D=4\times10^{-3}$, $d_0/L$ is the non-dimensional droplet diameter at the
initial time, the forcing-scale Weber number $We \equiv \rho
L_f^{3}F_0/\sigma$, where $\sigma$ is the surface tension, the Cahn number
$Ch=\xi/L$, where $\xi$ is the interface width, is kept fixed at $Ch=0.0028$,
 $\langle d_p\rangle_t/L$ is the
steady-state droplet diameter non-dimensionalized with the box length $L$, the
dissipation scale $\eta=\left(\nu^{3}/\varepsilon\right)^{\frac{1}{4}}$, where
$\varepsilon$ is the fluid-energy dissipation rate ($\varepsilon(t)=\langle \nu
|\omega({\bm x},t)|^{2} \rangle_{{\bm x}}$), $E(t)=\langle |{\bm u}({\bm
x},t)|^{2} \rangle_{{\bm x}}$ is the fluid kinetic energy, and $Re_{\lambda}$
the Taylor-microscale Reynolds number.  In all cases $\langle \rangle_t$
denotes the average over time in the statistically steady state.}
\end{table*}
\section{III. RESULTS}
Our investigations of droplet dynamics are divided
into two broad categories. We first elucidate the turbulence-induced
modification of the droplet 
in 
subsections A and B. 
Then we show how the droplet modifies various statistical properties
of turbulence, such as $E(k)$, in subsection C.
\subsection{A. Droplet deformation statistics}
\begin{figure*}
\includegraphics[width=.45\linewidth]{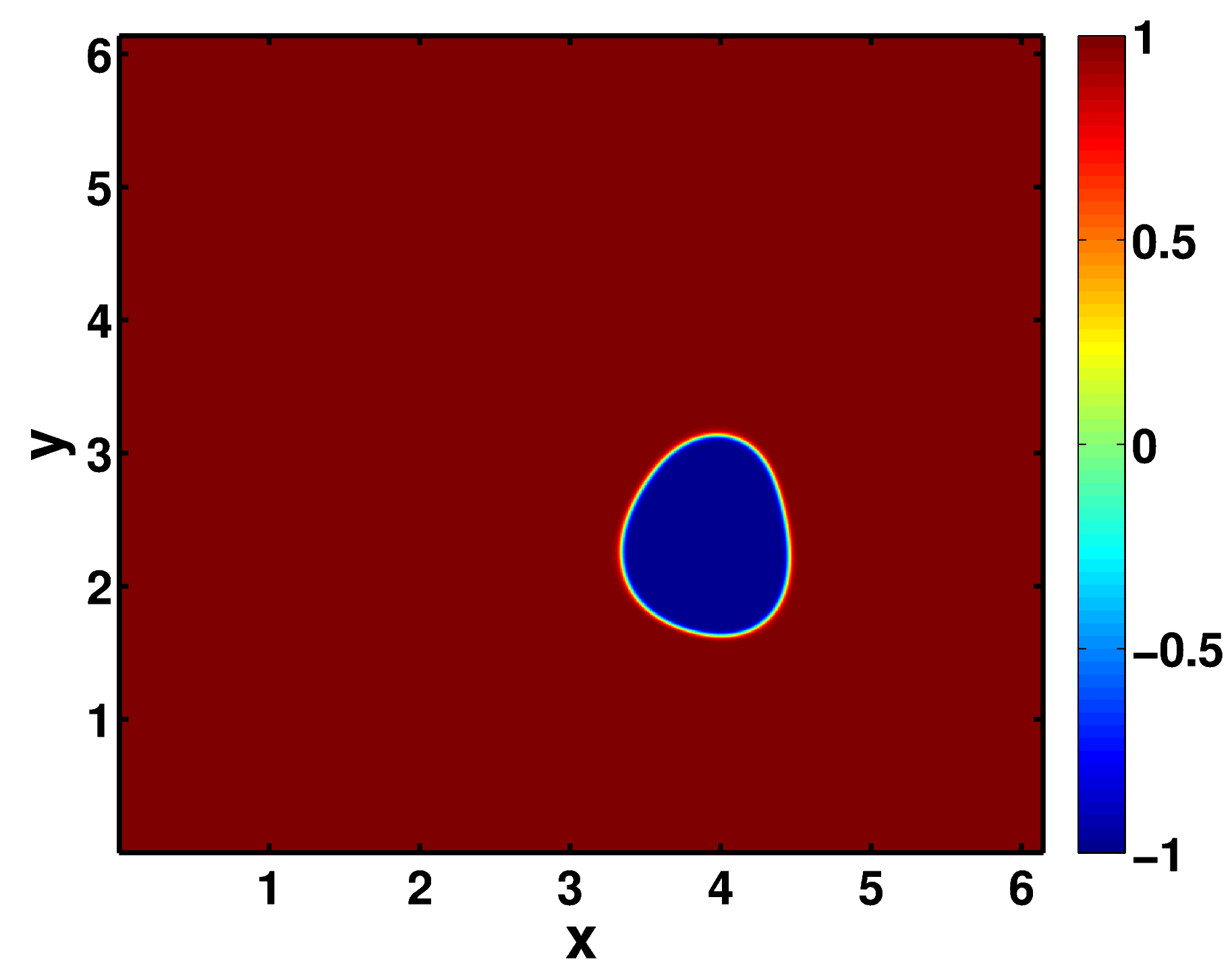}
\put(-185,160){\bf (a)}
\includegraphics[width=0.45\linewidth]{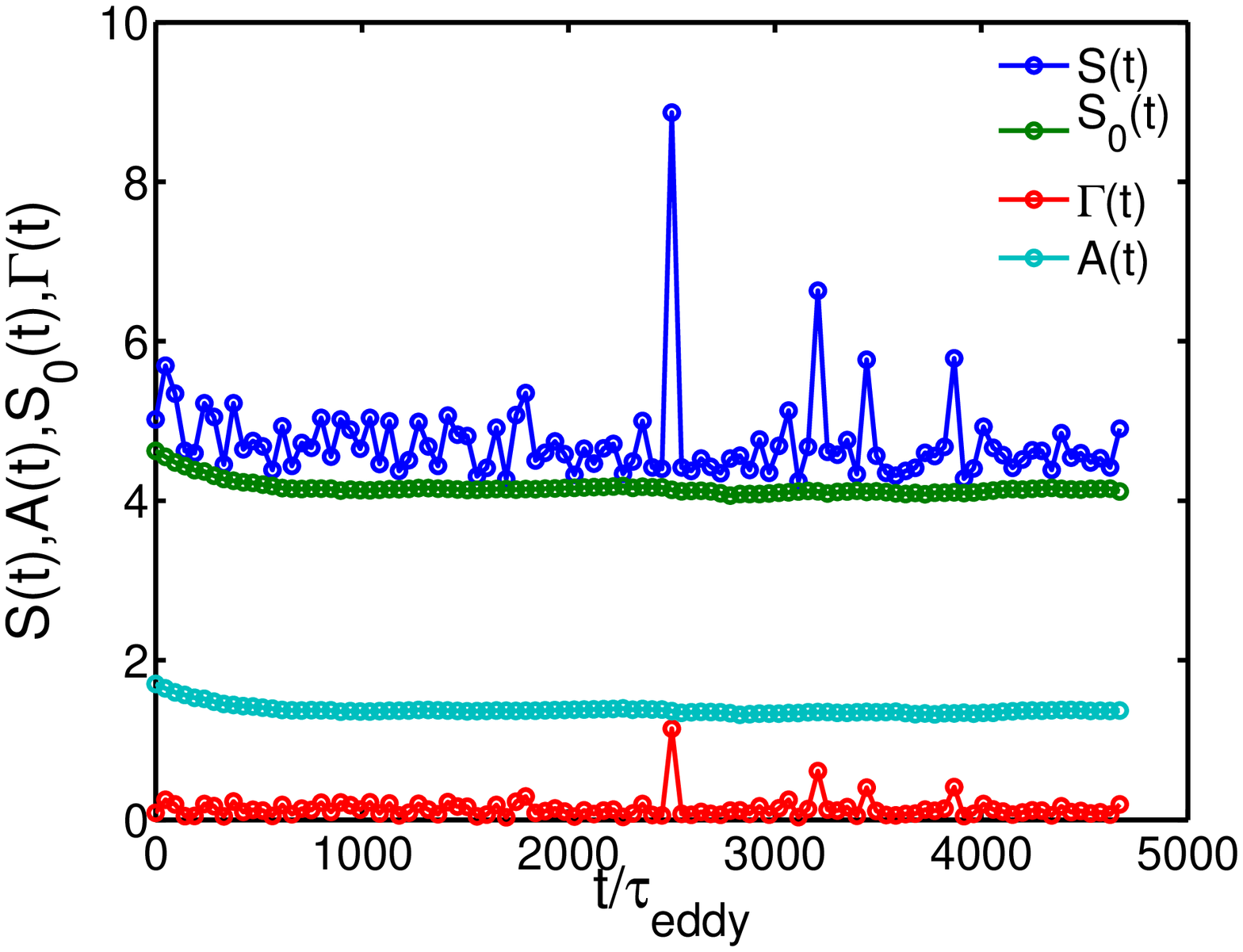}
\put(-185,160){\bf (b)}

\caption{(Color online) (a)Pseudocolor plot of the $\phi$ field;
(b) plots versus $t/\tau_{eddy}$ of the perimeter ${\mathcal {S}}(t)$ 
(deep-blue line), area $A(t)$ 
(light-blue line), perimeter ${\mathcal {S}}_0(t)$ 
(green line), of a circular droplet of area $A(t)$, and the deformation parameter 
$\Gamma(t)$ (red line) for the run {\tt{R7}} ($We=5.34$).}
\label{def}
\end{figure*}

We use Eq.~(\ref{peri}) for $\Gamma(t)$ and
 obtain $\mathcal{S}(t)$ by finding the length of the $\phi=0$ contour and 
the area $A(t)$
inside the $\phi=0$ contour. 
We then calculate $d_p(t)=2\sqrt{A(t)/\pi}$, an effective diameter for the droplet
that is not circular in general. Given the initial profile~(\ref{tanh}), 
we find that $\langle d_p \rangle_t < d_0$,
and $\langle d_p \rangle_t$ increases roughly 
linearly with $d_0$.
In Fig.~\ref{def}(b) we plot the perimeter ${\mathcal {S}}(t)$ (deep-blue line), area
$A(t)$ (light-blue line), the perimeter ${\mathcal {S}}_0(t)$ of a circular droplet of area
$A$  (green line), and the deformation parameter $\Gamma(t)$ (red line) for the
run {\tt{R7}} with $We=5.34$.  This plot shows that the instantaneous total
area $A(t)$ of the minority phase decreases very little over the entire
duration of our simulation. $A(t)$ is almost constant and 
just fluctuates about its mean value $\langle A(t) \rangle_t$; 
these fluctuations do not contribute significantly
to the deformation statistics because they are much smaller than the
fluctuations in the droplet perimeter ${\mathcal {S}}(t)$. (We expect that, in the limit of
zero mobility and constant surface tension (i.e., the sharp-interface limit),
the mass transfer is negligible, and $A(t)$ is independent of $t$.)

\begin{figure*}
\includegraphics[width=.33\linewidth]{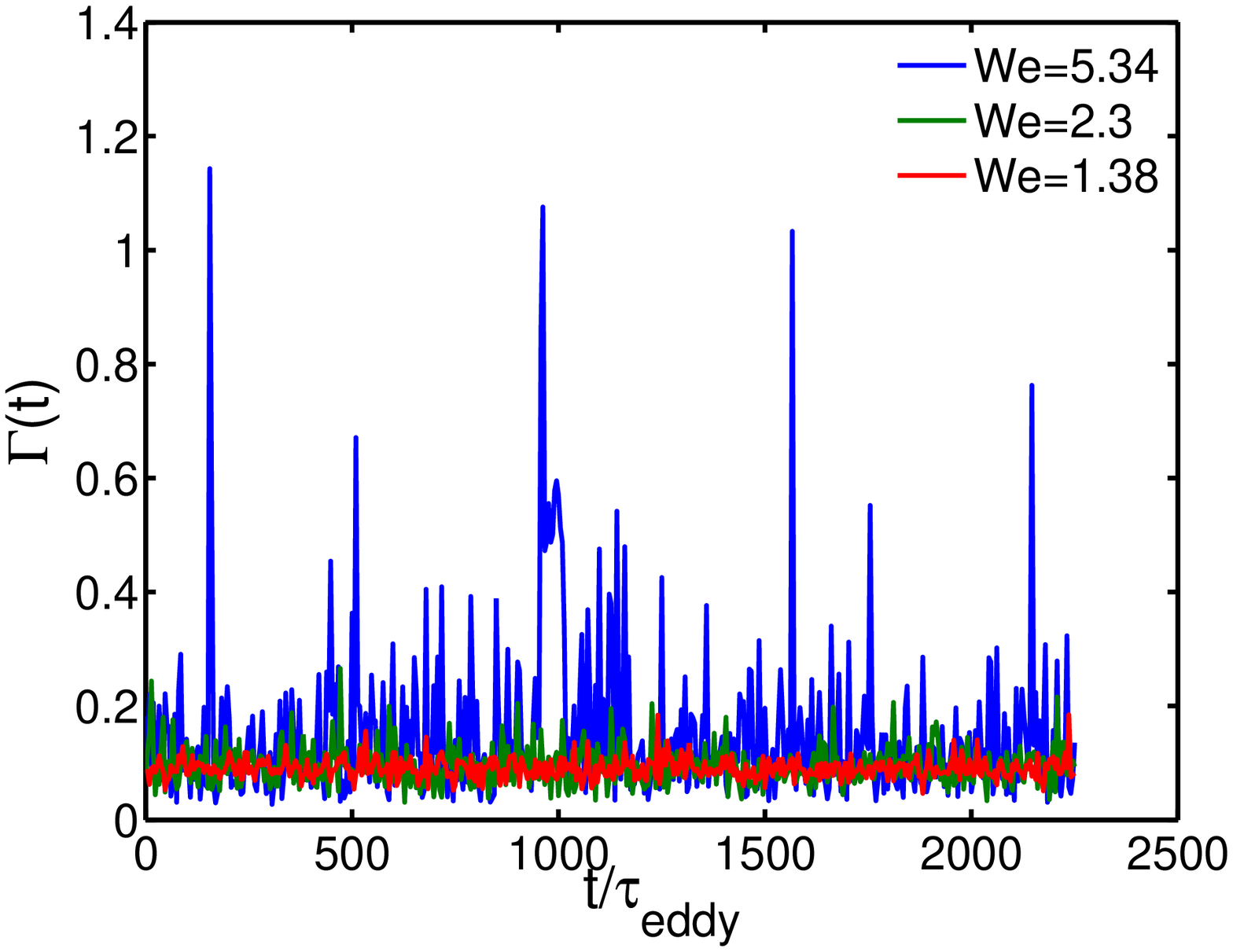}
\put(-140,110){\bf (a)}
\includegraphics[width=.33\linewidth]{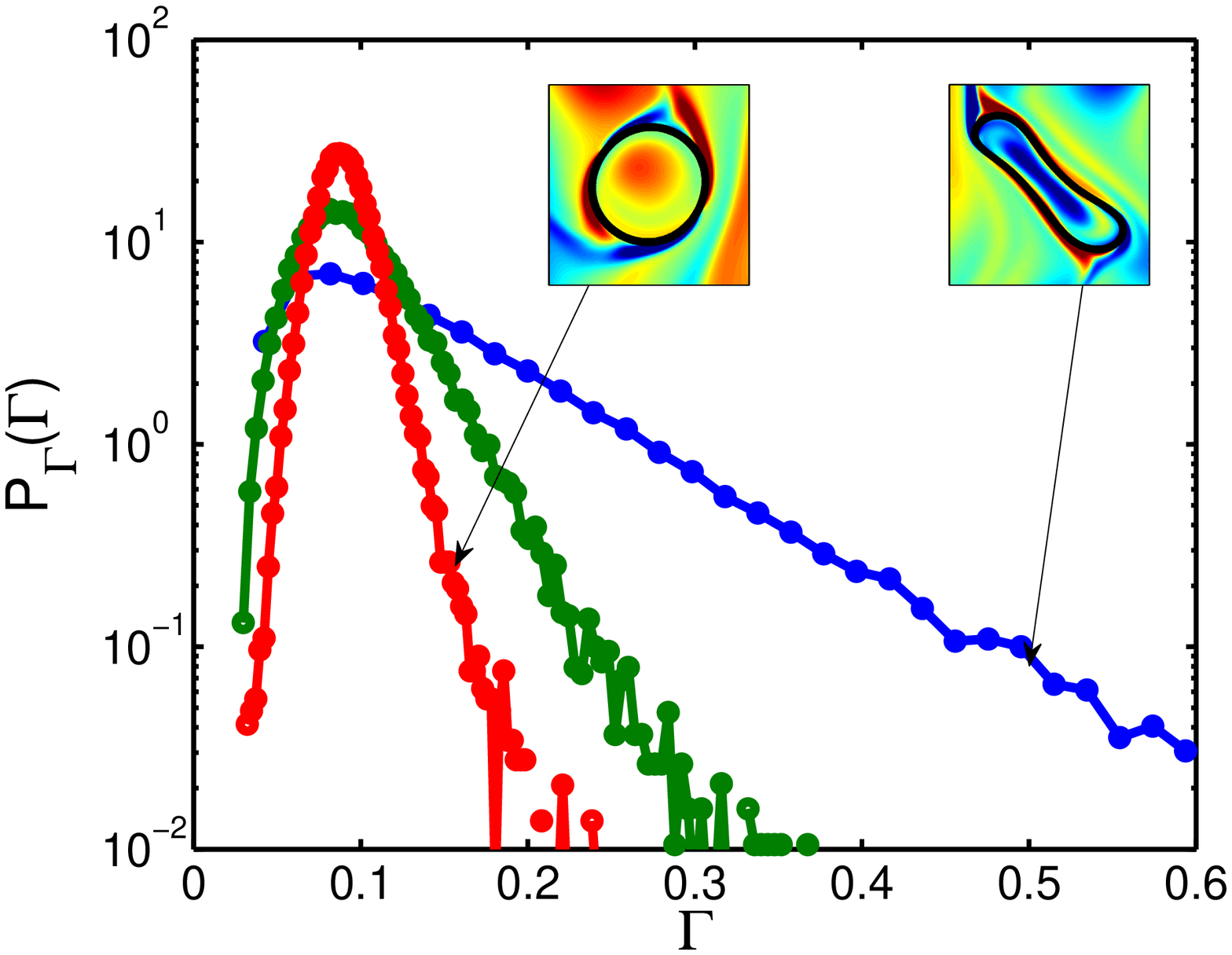}
\put(-140,110){\bf (b)}
\includegraphics[width=.33\linewidth]{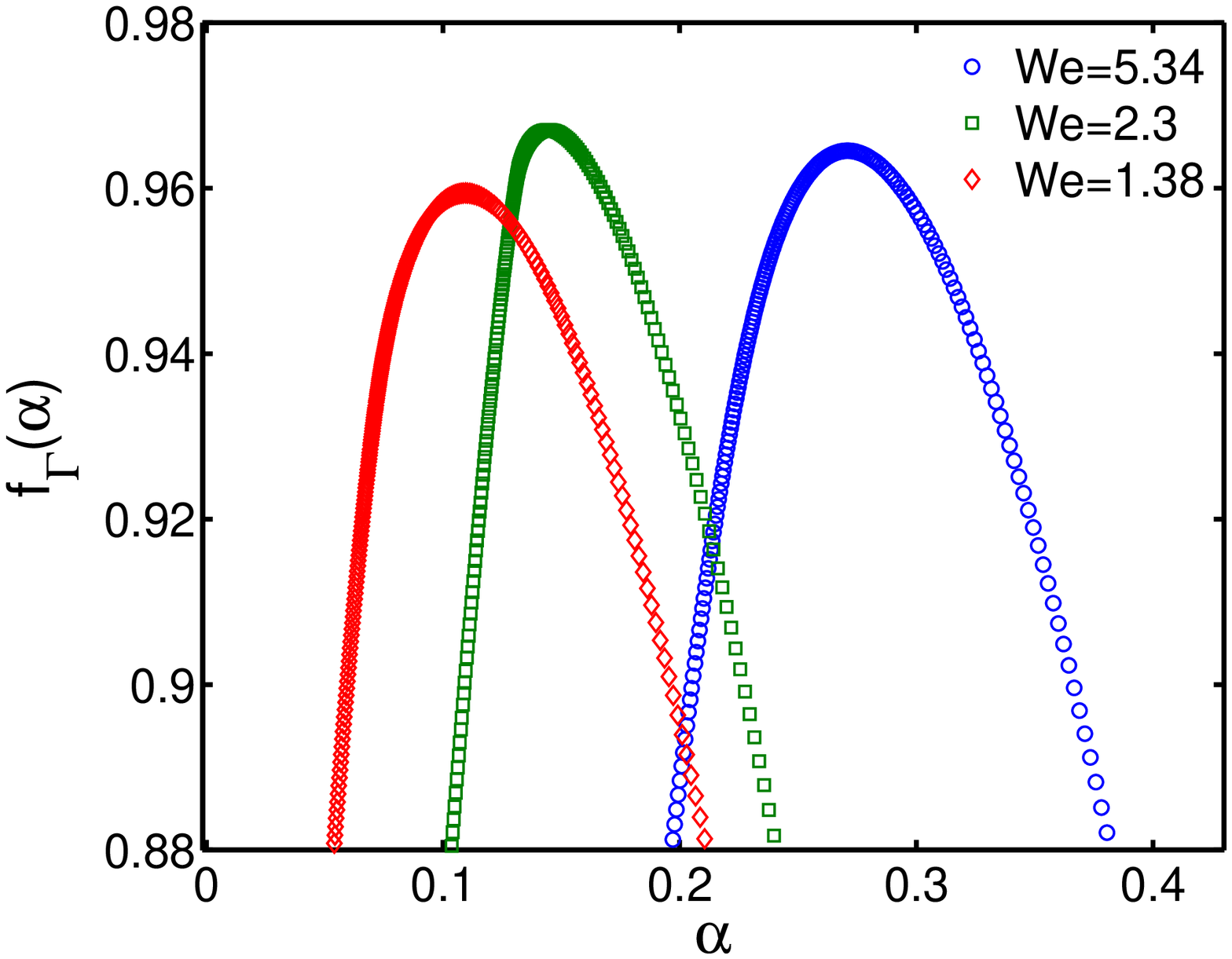}
\put(-140,110){\bf (c)}
\caption{(Color online) (a) Plots versus $t/\tau_{eddy}$ of $\Gamma(t)$
for the runs {\tt{R7}} ($We=5.34$, blue line), {\tt{R8}}
($We=2.3$, green line) and {\tt{R12}} ($We=1.38$, red line);
(b) plots of the PDFs $P(\Gamma)$, 
for the runs {\tt{R7}} ($We=5.34$, blue line with circles), {\tt{R8}} ($We=2.3$, green line with circles) 
and {\tt{R13}} ($We=1.38$, red line with circles);  (c) the multifractal
spectra $f_{\Gamma}(\alpha)$ for the timeseries of $\Gamma$ 
for the runs {\tt{R7}} ($We=5.34$, blue circles), {\tt{R8}} ($We=2.3$, green squares) 
and {\tt{R13}} ($We=1.38$, red diamonds).  The insets in (a)
show pseudocolor plots of the vorticity field with $\phi$-field contours
superimposed on them; the time evolution of such plots are given in the videos
{\tt V1} and {\tt V2} in Ref.~\cite{suppmat}}
\label{deformation}
\end{figure*}

Our droplet diameters are comparable to lengths in the inertial range, which
lies in between the large forcing length scale and the small scales where
dissipation is significant. Turbulence induces large fluctuations in the shape
of a droplet, so we integrate Eqs.(\ref{ns}) and (\ref{ch}) for $2000
\tau_{eddy}$, to obtain the 
time series of the dimensionless deformation $\Gamma(t)$, which we depict in
Figs.~\ref{deformation}(a), for different values of $We$. 
Not
only does the mean $\langle \Gamma \rangle_{t}$ increase as $We$ increases, so
do the variance, skewness, and kurtosis of this time series. 
In particular, the
root-mean-square value
$\Gamma_{rms}=\langle\left(\Gamma-\langle\Gamma\rangle_t\right)^{2}\rangle_t$ increases
with $We$ ($\Gamma_{rms}=0.14$ for $We=5.34$, $\Gamma_{rms}=0.033$ for $We=2.3$ and $\Gamma_{rms}=0.016$ for $We=1.38$),
as do the
 skewness $\gamma_1 =
\langle\left(\Gamma-\langle\Gamma\rangle_t\right)^{3}\rangle_t/\langle\left(\Gamma-\langle\Gamma\rangle_t\right)^{2}\rangle_t^{3/2}$
($\gamma_1=2.9$ for $We=5.34$, $\gamma_1=1.57$ for $We=2.3$ and $\gamma_1=0.8$ for $We=1.38$) and the kurtosis
$\gamma_2=\langle\left(\Gamma-\langle\Gamma\rangle_t\right)^{4}\rangle_t/\langle\left(\Gamma-\langle\Gamma\rangle_t\right)^{2}\rangle_t^{2}$
($\gamma_2=22.4$ for $We=5.34$, $\gamma_2=7.5$ for $We=2.3$ and $\gamma_2=5.8$ for $We=1.38$).
We find that $\Gamma_{rms},\gamma_1$, and $\gamma_2$ decrease as
$We$ decreases (i.e., the surface tension $\sigma$
increases) and the droplet becomes rigid.

From the time series of $\Gamma(t)$ we find the PDF $P_{\Gamma}(\Gamma)$
(Fig.~\ref{deformation}(b)).
These plots quantify
the intuitively appealing result that the fluctuations of the droplet increase
with an increase in $We$ (i.e., decrease with an increase in $\sigma$). 
The right tail of $P_{\Gamma}(\Gamma)$
decays exponentially with $\Gamma$; this decay steepens as $We$ decreases, and
$P_{\Gamma}(\Gamma)$ sharpens, as it must, for there can be no shape
fluctuations if $We=0$ (a perfectly rigid droplet).

The time series of $\Gamma(t)$ and the large kurtosis of $P_{\Gamma}(\Gamma)$
 suggest intermittency; we
characterize this intermittency by obtaining the multifractal spectrum~(see
Refs.\cite{muzy1993,ashken,physionet}) $f_{\Gamma}(\alpha)$ (Fig.~\ref{deformation}(c)),
which is the Legendre transform of the Renyi exponents $\tau(q)$ that follow
from $\langle |\Gamma(0)-\Gamma(t)|\rangle^{q}\sim t^{\tau(q)}$.  This
remarkable multifractality of $\Gamma(t)$ has not been
noted so far. As $We$
decreases ($\sigma$ increases), the droplet-shape fluctuations decrease and the
value of $\alpha$, at which $f_{\Gamma}(\alpha)$ attains a maximum, shifts
towards $0$.  If $\sigma$ is low, the droplet can break up at certain times,
but the broken fragments coalesce to form a single drop again.  The break-up
events can be identified from the largest spikes in $\Gamma(t)$, because the
formation of small droplets increases the total perimeter. Such droplet
breakups occur only with the smallest value of $\sigma$ that we consider, and
then only for about $4\%$ of the total time.
We give an outline of the method we use to obtain multifractal spectra in the Appendix,
where we follow Refs.~\cite{muzy1993,ashken,physionet}.

\subsection{B. Droplet center-of-mass acceleration statistics}
\begin{figure*}
\includegraphics[width=.45\linewidth]{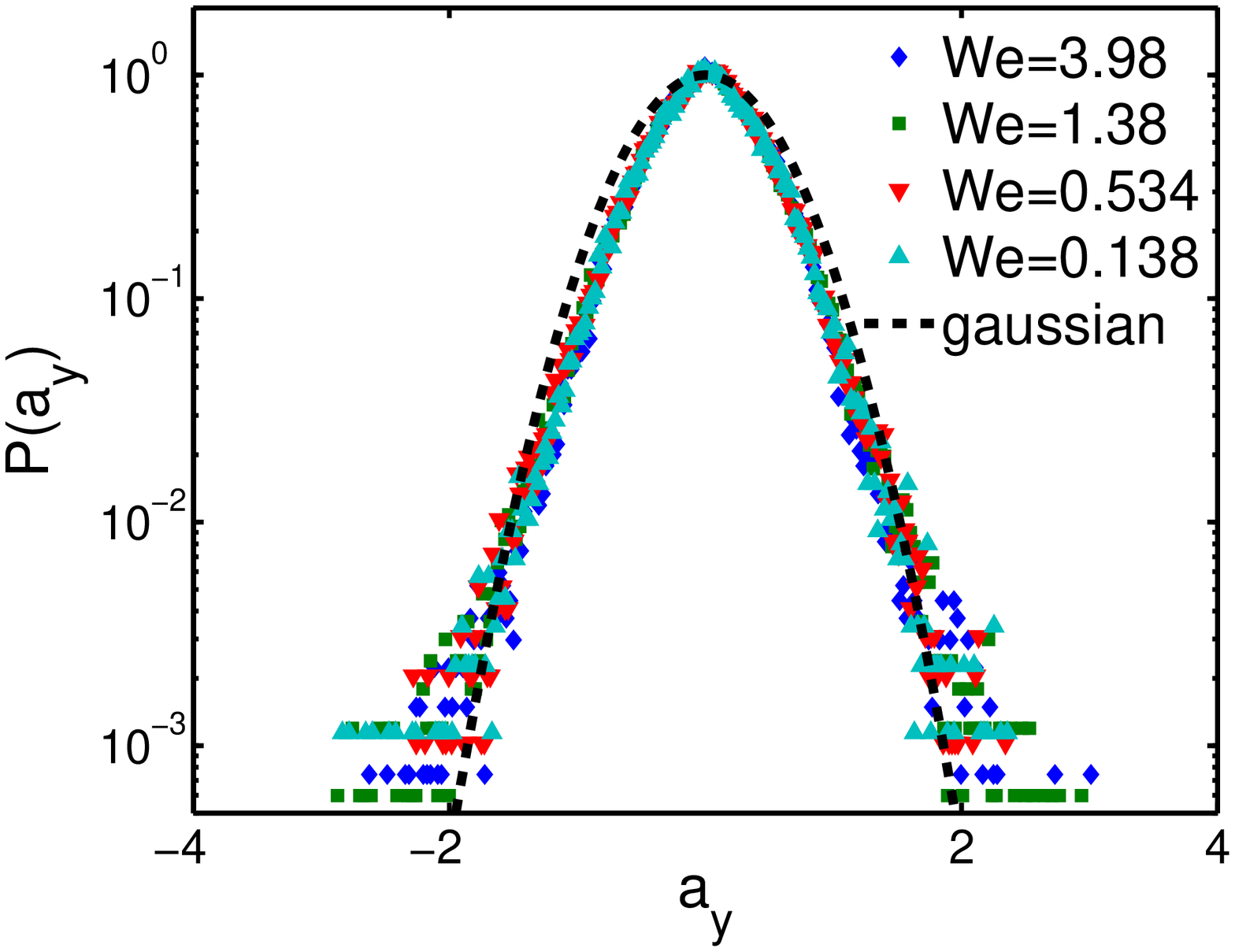}
\put(-185,160){\bf (a)}
\includegraphics[width=.45\linewidth]{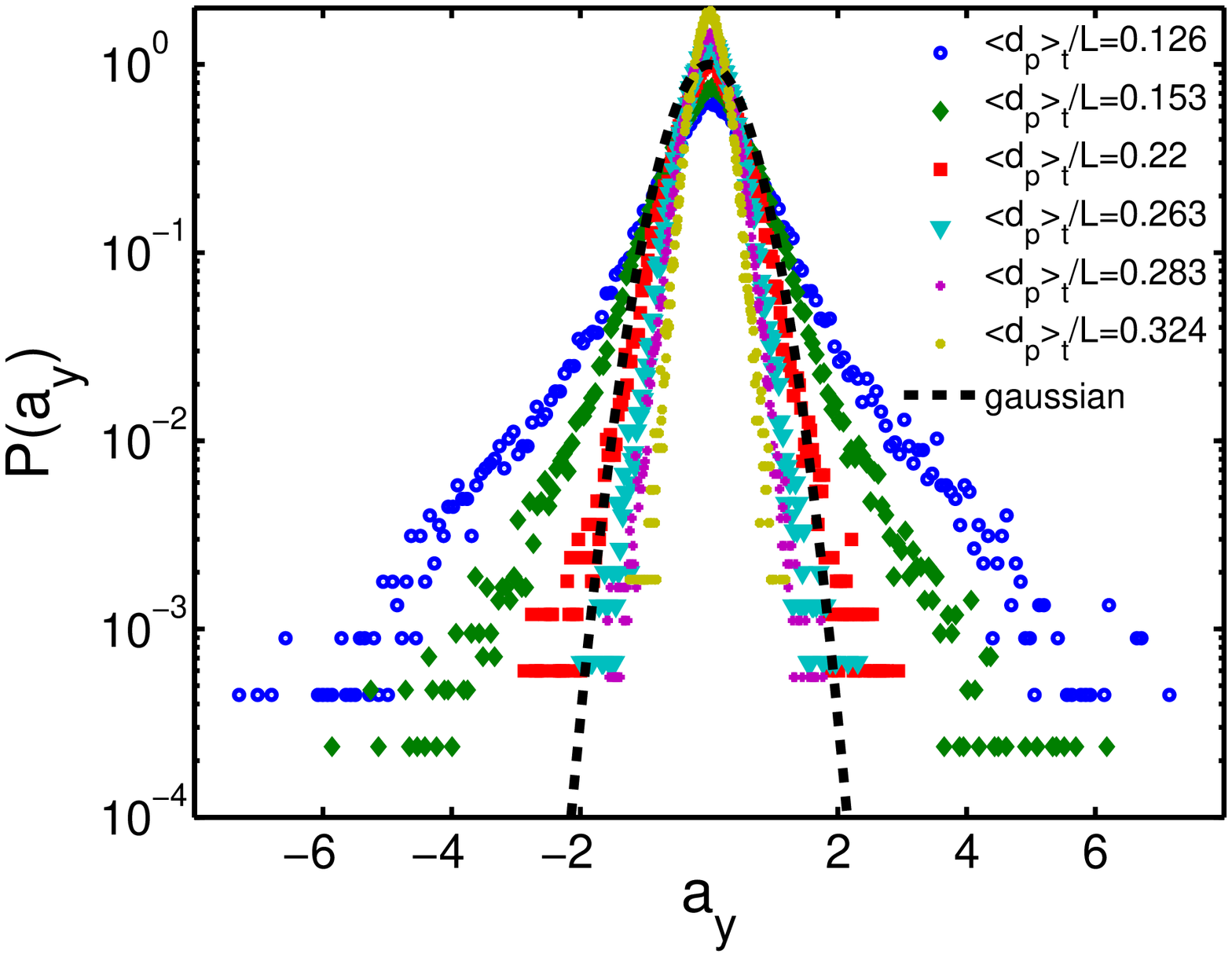}
\put(-185,160){\bf (b)}

\includegraphics[width=.45\linewidth]{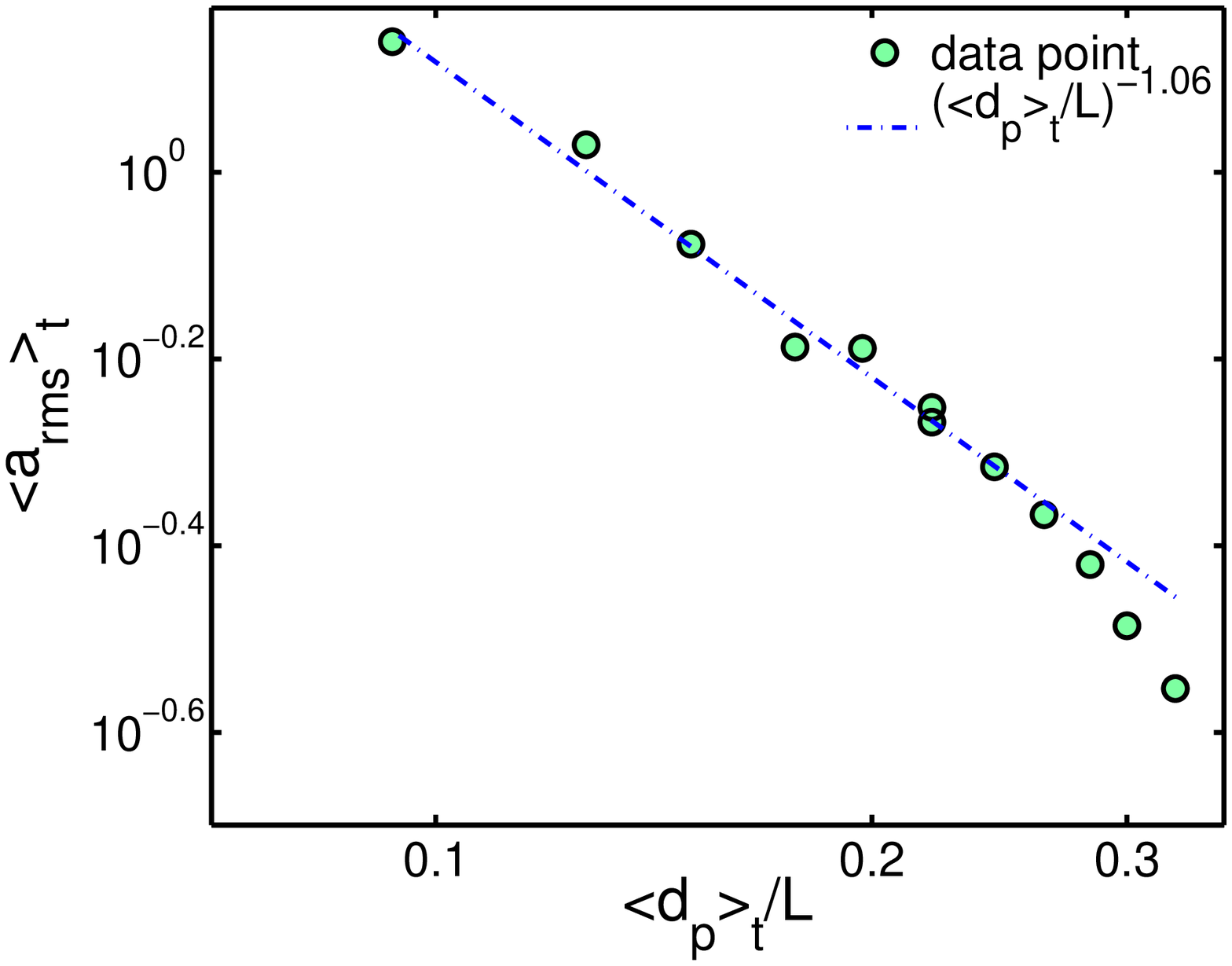}
\put(-185,160){\bf (c)}
\includegraphics[width=.45\linewidth]{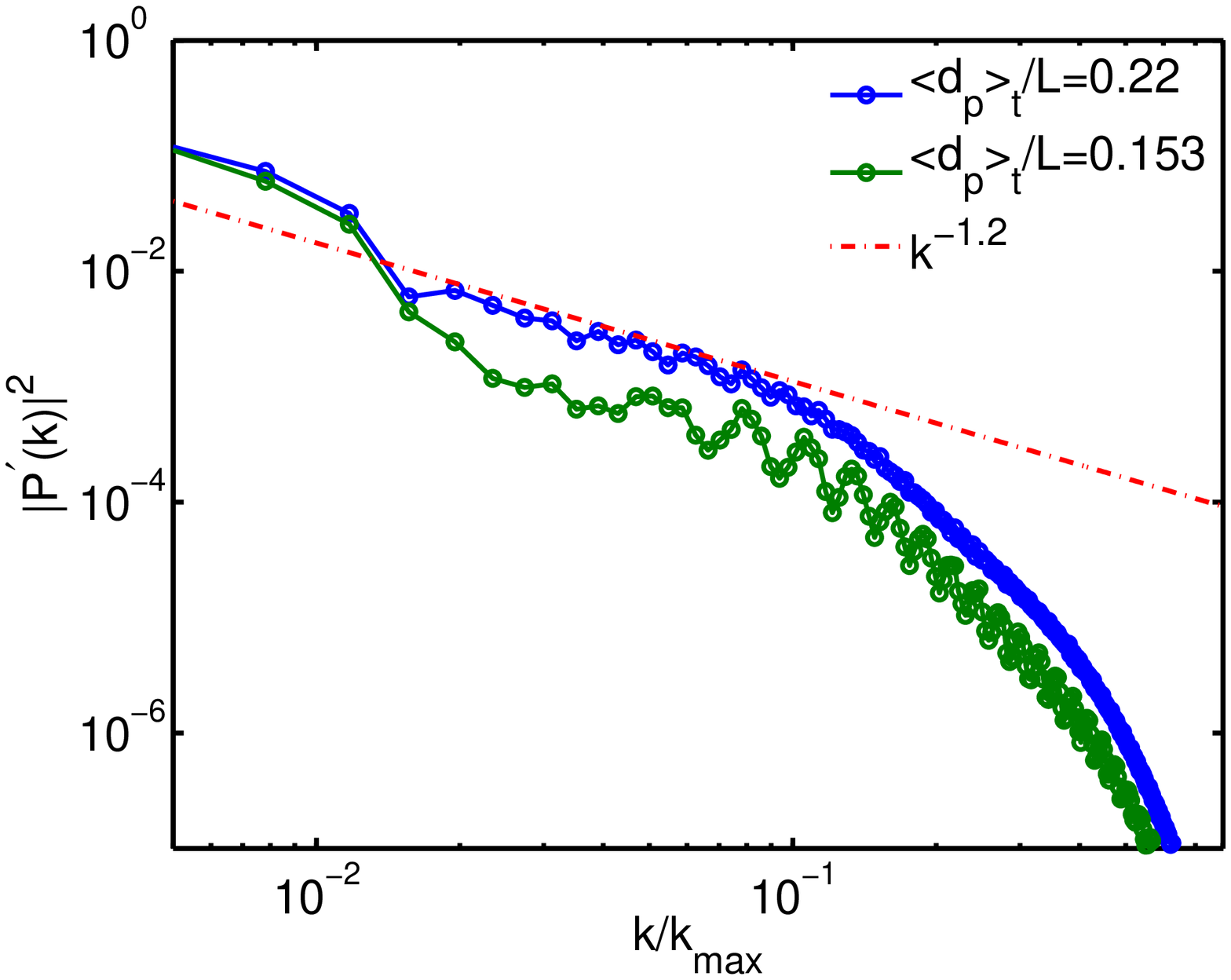}
\put(-165,160){\bf (d)}
\caption {(Color online) (a) Semilog (base $10$) plots at $Gr=3 \times 10^{7}$ of $P(a_y)$, the PDF of
$a_{y}$ of the center of mass of the droplets, for runs {\tt{R8}}
($We=2.3$, deep-blue diamonds), {\tt{R12}} ($We=1.38$, green squares),
{\tt{R13}} ($We=0.534$, red inverted triangles) and {\tt{R14}}
($We=0.138$, light-blue triangles), at $\langle d_p
\rangle_t/L=0.22$; (b) {\tt{R20}} ($\langle d_p\rangle_t/L=0.126$, deep-blue
circles), {\tt{R17}} ($\langle d_p\rangle_t/L=0.153$, green diamonds),
{\tt{R12}} ($\langle d_p\rangle_t/L=0.22$, red squares), {\tt{R5}} ($\langle
d_p\rangle_t/L=0.263$, light-blue inverted triangles), {\tt{R4}}
($\langle d_p\rangle_t/L=0.283$, magenta plus signs) and {\tt{R2}} ($\langle
d_p\rangle_t/L=0.324$, yellow asterisk) at $We=1.38$; (c) plot
of $\langle a_{rms}\rangle_t$ versus $\langle d_p\rangle_t/L$; (d) Log-log plots (base $10$) versus the scaled
wavenumber $k/k_{max}$ of the pressure spectrum $|\mathcal{P}(k)|^{2}$
for runs {\tt{R12}} ($\langle d_p \rangle_t/L=0.22$, deep-blue line with circles), {\tt{R17}} ($\langle d_p \rangle_t/L=0.177$, green line with circles), 
{\tt{R1}} (single-phase fluid, red
line with circles), power-law scaling $k^{-1.2}$ (light-blue and magenta dash-dot line) and $k^{-9}$
(yellow dash-dot line).
In (a) and (b)
the black dashed line shows a Gaussian fit. }
\label{spectra}
\end{figure*}

We now investigate
the advection of the droplet inside the background fluid.
To quantify droplet advection, we obtain PDFs of the components of the
acceleration of the center of mass of the droplet along its
trajectory~\cite{biferale13conf}. We obtain the
center of mass velocity ${\bf v}_{CM}$ of the droplet and
$a_y$, the $y$
component of the acceleration of the droplet center of mass,
where 
\begin{eqnarray}
{\bf v}_{CM}(t) & = & \sum \limits _{{\bm x} \ni \phi({\bm x},t)<0} {\bm u} ({\bm x},t)\\
\label{velcm}
{\rm and} \qquad a_y(t) & = & \sum \limits_{{\bm x} \ni \phi({\bm x},t)<0} (Du_y({\bm x},t)/Dt).
\label{acc}
\end{eqnarray}
Note that $\phi({\bm x}, t) < 0$ if ${\bm x}$ lies inside the droplet at 
time $t$, and
$D/Dt=\partial_t + {\bm u}\cdot \nabla$.
We present results for $a_y$ (the results for the $x$ component $a_x$ are similar),
and the root-mean-square acceleration
$a_{rms}=\sqrt{a_{y}^{2}+a_x^{2}}$. We restrict ourselves to values of $\sigma$
for which there is a single droplet in the flow; and we use $10$ different
values of $d_0$ in the range $0.134L$ to $0.334L$.  In Fig.~\ref{spectra}(a) we
plot the PDF $P(a_y)$ for four different values of $We$ at $d_0/L=0.24$. These
PDFs collapse on top of each other (Fig.~\ref{spectra}(a)), so, in a
statistical sense, the center of mass of a deformable droplet moves in the same
way as a rigid droplet.  Indeed, $P(a_y)$ is very close to a Gaussian (black
dashed line), for droplets with $d_0/L=0.24$.  
From Eq.~(\ref{acc}) we see that the acceleration of the center
of mass of the droplet follows from an integral over the area of the droplet.
For a rigid droplet, whose diameter is comparable to inertial-range scales, we
expect the small-scale fluctuations to be averaged out and $P(a_y)$ to be close
to a Gaussian. We do, indeed, find this, for several values of $We$,  in
Fig.~\ref{spectra}(a), where $\langle d_p \rangle_t/L = 0.22$. By contrast,
when we reduce $\langle d_p \rangle_t/L$, this PDF shows  significant
deviations from a Gaussian form as we show in Fig.~\ref{spectra}(b).  

Our
results for $P(a_y)$ are in qualitative accord with those for the advection of
a rigid particle by a three-dimensional (3D), homogeneous and isotropic
turbulent flow~\cite{jeremie}, for particle diameters in the inertial range:
References~\cite{jeremie, qureshi2007} suggest that plots of the velocity
variance $|\frac{|{\bf v}_{CM}|^{2}-u_{rms}^{2}}{u_{rms}^{2}}|$, $\langle
a_{y}^{2} \rangle$, and $\langle a_{rms} \rangle_t$ versus the scaled particle
diameter $\left(\langle d_p \rangle_t/L\right)$ should exhibit power laws with
exponents that can be related to the inertial-range, power-law exponent in the
pressure spectrum. We adapt these arguments to our study of a droplet, with
mean scaled diameter $\langle d_p \rangle_t/L$. The plot in
Fig.~\ref{spectra}(c) is consistent with a power-law dependence of $\langle
a_{rms}\rangle_t$ on $\langle d_p \rangle_t/L$, albeit over a small
range~\cite{jeremie_footnote}, with exponents that can be related to the
inertial-range scaling of the pressure spectrum. If the pressure spectrum of
the turbulent fluid with a droplet is $|\tilde{\mathcal{P}}(k)|^{2}\sim
k^{-\alpha_{\mathcal{P}}}$, for $k$ in the scaling range, then $\langle a_{rms}
\rangle_t \sim \left(\langle
d_p\rangle_t/L\right)^{\frac{\alpha_{\mathcal{P}}-3}{2}}$.
We give details of the relation between the pressure-spectrum scaling
and the plot of the acceleration variance versus the non-dimensionalized droplet diameter
scaling below.

Our simulations suggest that  
$\langle a_{rms} \rangle \sim \left(\langle d_p\rangle_t/L\right)^{-1.06}$.  
Here we provide arguments that suggest
such a power-law dependence; we follow the treatment of Refs.~\cite{jeremie,
qureshi2007} for rigid particles.  
We first define the structure function for increments of the pressure $\mathcal{P}$ as
\begin{equation}
S_{2}^{{\mathcal{P}}}({\mathbf{r}})  =  \langle \left({\mathcal{P}}({\mathbf{x}})-{\mathcal{P}}({\mathbf{x}}+{\mathbf{r}})\right)^{2} \rangle
\sim r^{\zeta_{2}^{{\mathcal{P}}}},
\end{equation}
for separations $r$ in the inertial range. If we introduce
$\tilde{{\mathcal{P}}}({\mathbf{k}}) = (1/4\pi^{2})\int d {\mathbf{x}} e^{i{\mathbf{x.k}}} {\mathcal{P}}({\mathbf{x}})$, the spatial Fourier transform
of ${\mathcal{P}}({\mathbf{r}})$, we have
\begin{eqnarray}
S_{2}^{{\mathcal{P}}}({\mathbf{r}}) & = & \langle {\mathcal{P}}({\bf x}+{\bf r}) \rangle^{2} + 
\langle {\mathcal{P}}({\bf x}) \rangle^{2} - 2 \langle {\mathcal{P}}({\bf x} + {\bf r}){\mathcal{P}}({\bf x}) \rangle, \nonumber \\ 
&=& 2 \int \limits_{0}^{\infty}dk |\tilde{\mathcal{P}}(k)|^{2} - \int \limits_{0}^{\infty} dk |\tilde{\mathcal{P}}(k)|^{2} \int \limits_{0}^{2\pi}e^{-ikrcos\theta}d\theta, \nonumber \\
&=& 2 \int \limits_{0}^{\infty}dk |\tilde{\mathcal{P}}(k)|^{2}\left(1 - \pi I_{0}(r)\right) , 
\end{eqnarray}
where $I_{0}(r)=\sum \limits_{m=0} ^{\infty}\frac{1}{m!\Gamma(m+1)}\left(\frac{r}{2}\right)^{2m}$
is the modified Bessel function of the first kind. If we have the inertial-range scaling form $|\tilde{\mathcal{P}}(k)|^{2}\sim k^{-\alpha^p}$,
then the exponent
\begin{equation}
\alpha^{\mathcal{P}}=\zeta^{\mathcal{P}}_{2} + 1.
\end{equation}

In the velocity formulation of the NS equation
\begin{equation}
 \left(\partial_t + {\bm u}\cdot \nabla\right) {\bm u}= -\nabla {\mathcal{P}}/\rho + \nu \nabla^2 {\bm u} - \alpha {\bm u}-(\phi \nabla \mu) + F_{\bm u},
\label{ns1}
\end{equation}
we can assume that, \textit{in the inertial range}, the main contribution to the right-hand side of Eq.(~\ref{ns})
comes from (we take $\rho=1$) $-\nabla {\mathcal{P}}-(\phi \nabla \mu)\equiv-\nabla {\mathcal{P}}^{\prime}$.
We have introduced ${\mathcal{P}}^{\prime}$, so we now work with primed exponents $\alpha^{\mathcal{P}^{\prime}}$ and $\zeta_{2}^{\mathcal{P}^{\prime}}$,
which can be defined like their counterparts without the primes.  From Refs.~\cite{qureshi2007,hill} we know that
\begin{eqnarray}
\langle a_{rms}^{2}\rangle & \sim & \langle \left(\left(\partial_t + {\bm u}\cdot \nabla\right) {\bm u}\right)^{2} \rangle \nonumber \\
&\sim & \langle 
\nabla {\mathcal{P}}^{\prime} ({\bf{x}}+{\bf r})\nabla {\mathcal{P}}^{\prime}({\bf x})\rangle \nonumber \\
 & \sim & S_{2}^{\mathcal{P}^{\prime}}({\bf r})/r ,
\end{eqnarray}
so we have the scaling results
\begin{equation}
\langle a_{rms} \rangle \sim  \sqrt{S_2^{\mathcal{P}^{\prime}}({\mathbf{r}})}/r 
 \sim  r^{\zeta_{2}^{\mathcal{P}^{\prime}}/2}/r \sim r^{\frac{\alpha^{\mathcal{P}^{\prime}}-3}{2}} .
\end{equation}
From our simulations we find $\alpha^{\mathcal{P}^{\prime}}\simeq 1.2$ (Fig.~\ref{spectra}(d)), which implies $\langle a_{rms} \rangle \sim r^{-0.9}$,
which is consistent, given our error bars, with our measured value of $-1.06$ (Fig.~\ref{spectra}(b)); here $\langle d_p\rangle_t/L$ plays
the role of $r$ in our scaling arguments.


\subsection{C. Energy-dissipation time series and energy and order-parameter spectra}
\begin{figure*}
\includegraphics[width=.33\linewidth]{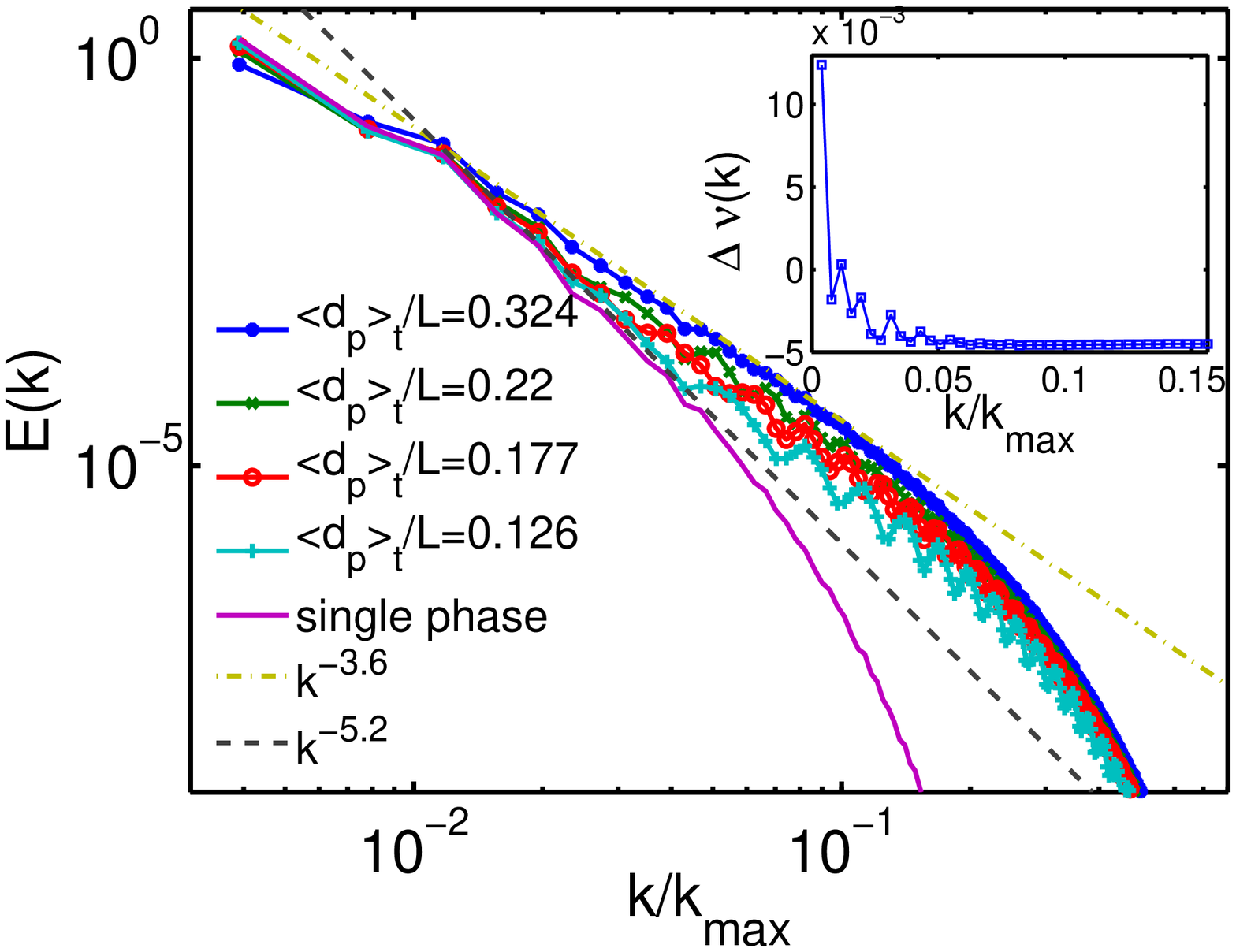}
\put(-135,110){\bf \scriptsize(a)}
\includegraphics[width=.33\linewidth]{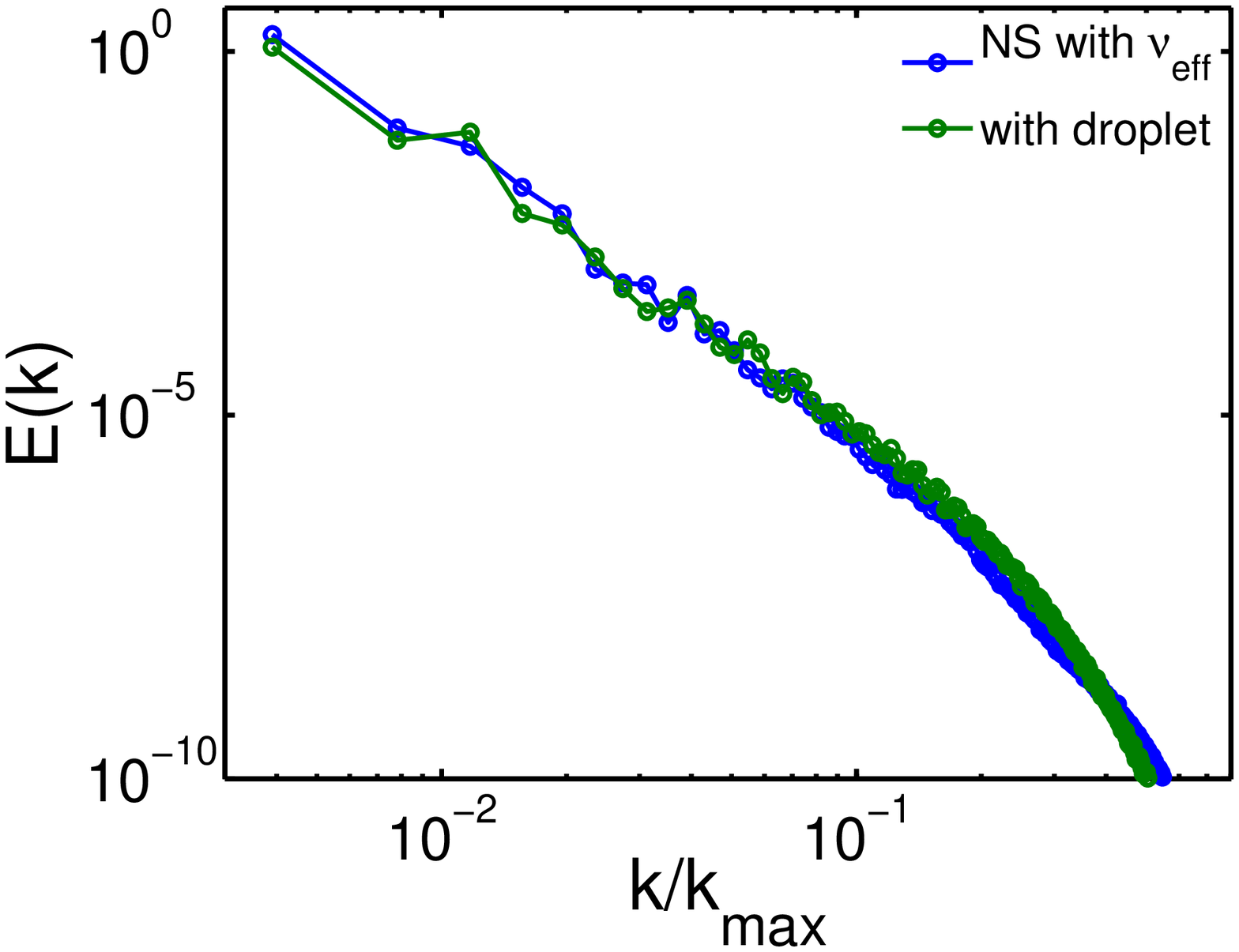}
\put(-135,110){\bf \scriptsize(b)}
\includegraphics[width=.33\linewidth]{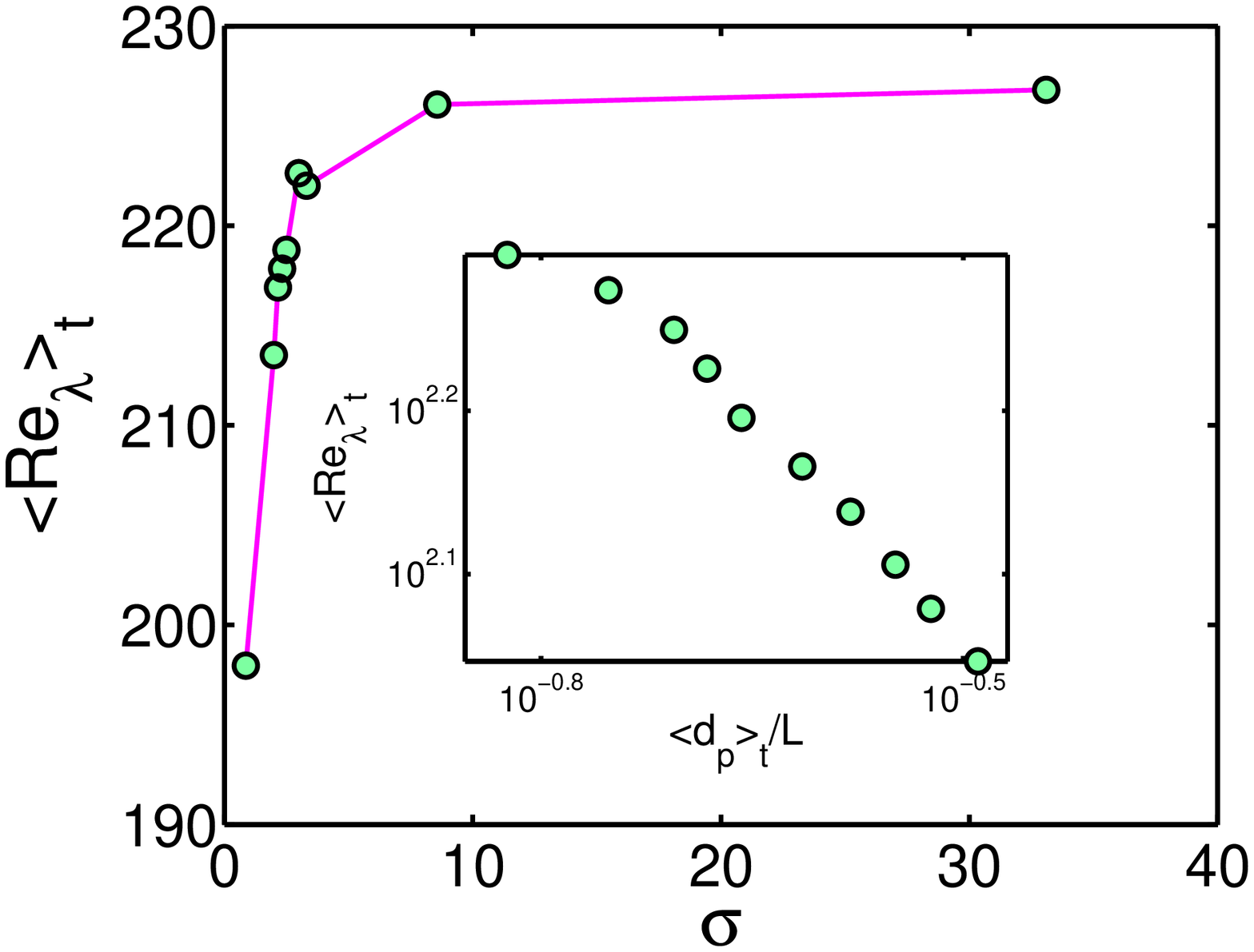}
\put(-135,110){\bf \scriptsize(c)}

\includegraphics[width=.33\linewidth]{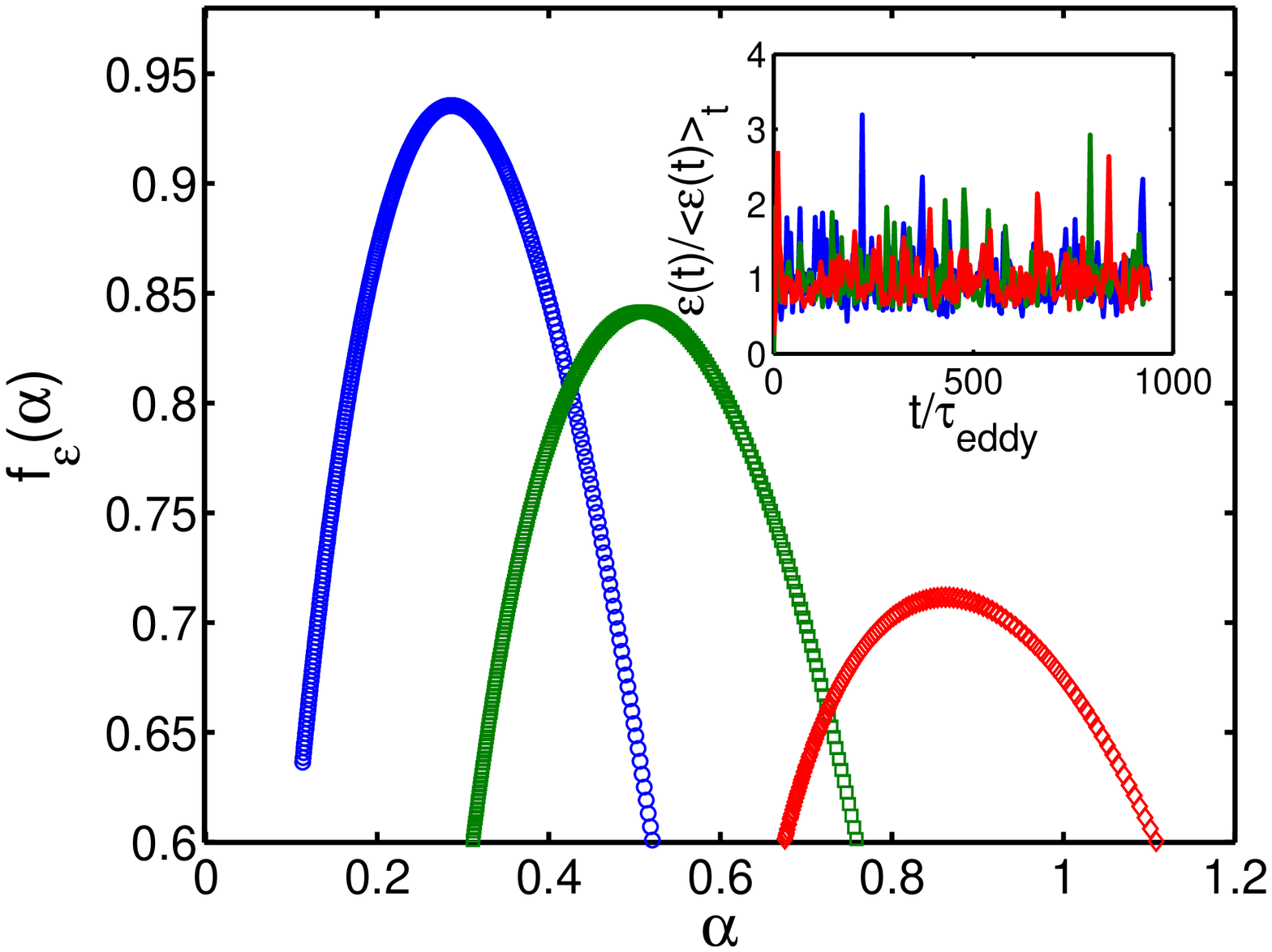}
\put(-135,110){\bf \scriptsize(d)}
\includegraphics[width=.33\linewidth]{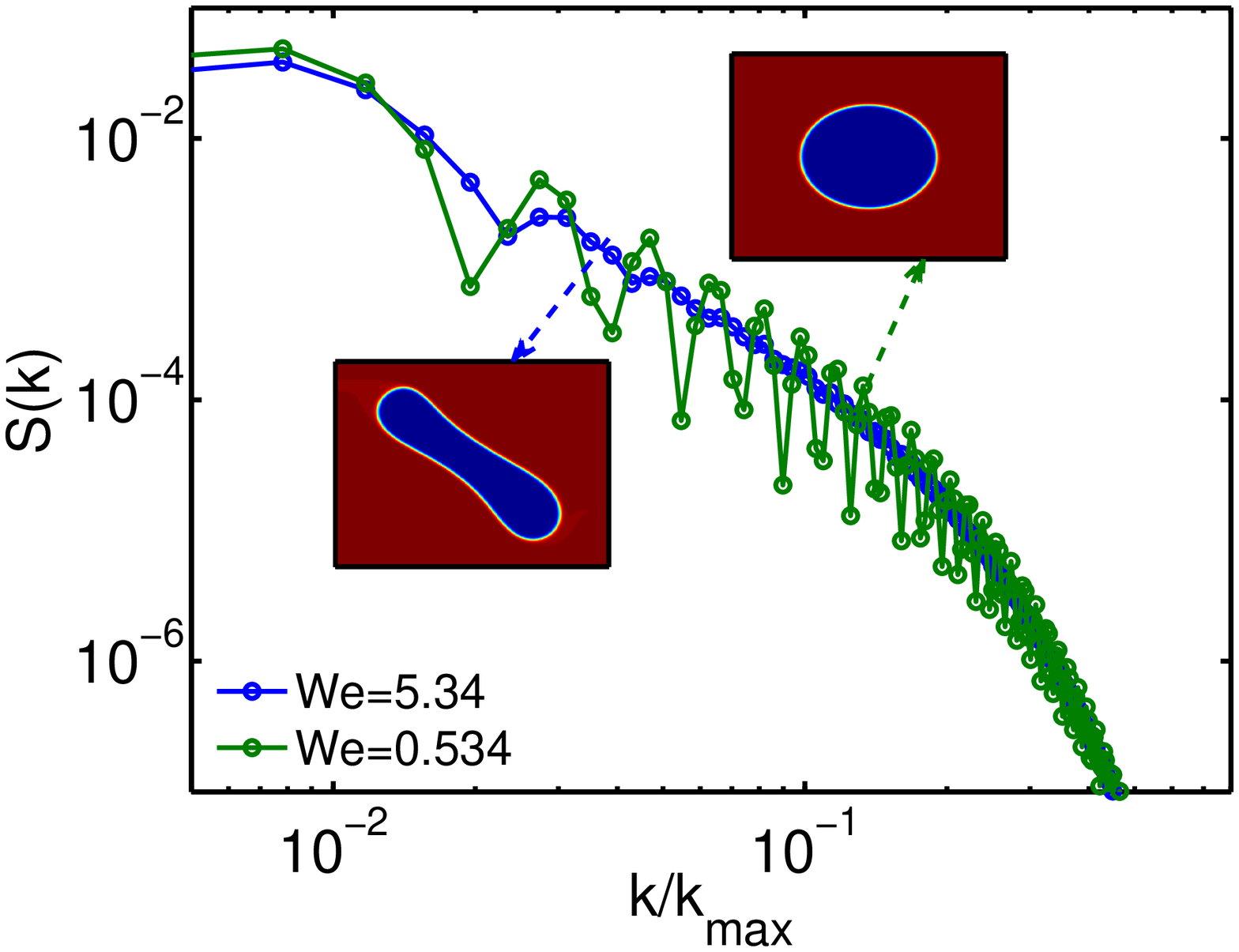}
\put(-135,110){\bf \scriptsize(e)}
\includegraphics[width=.33\linewidth]{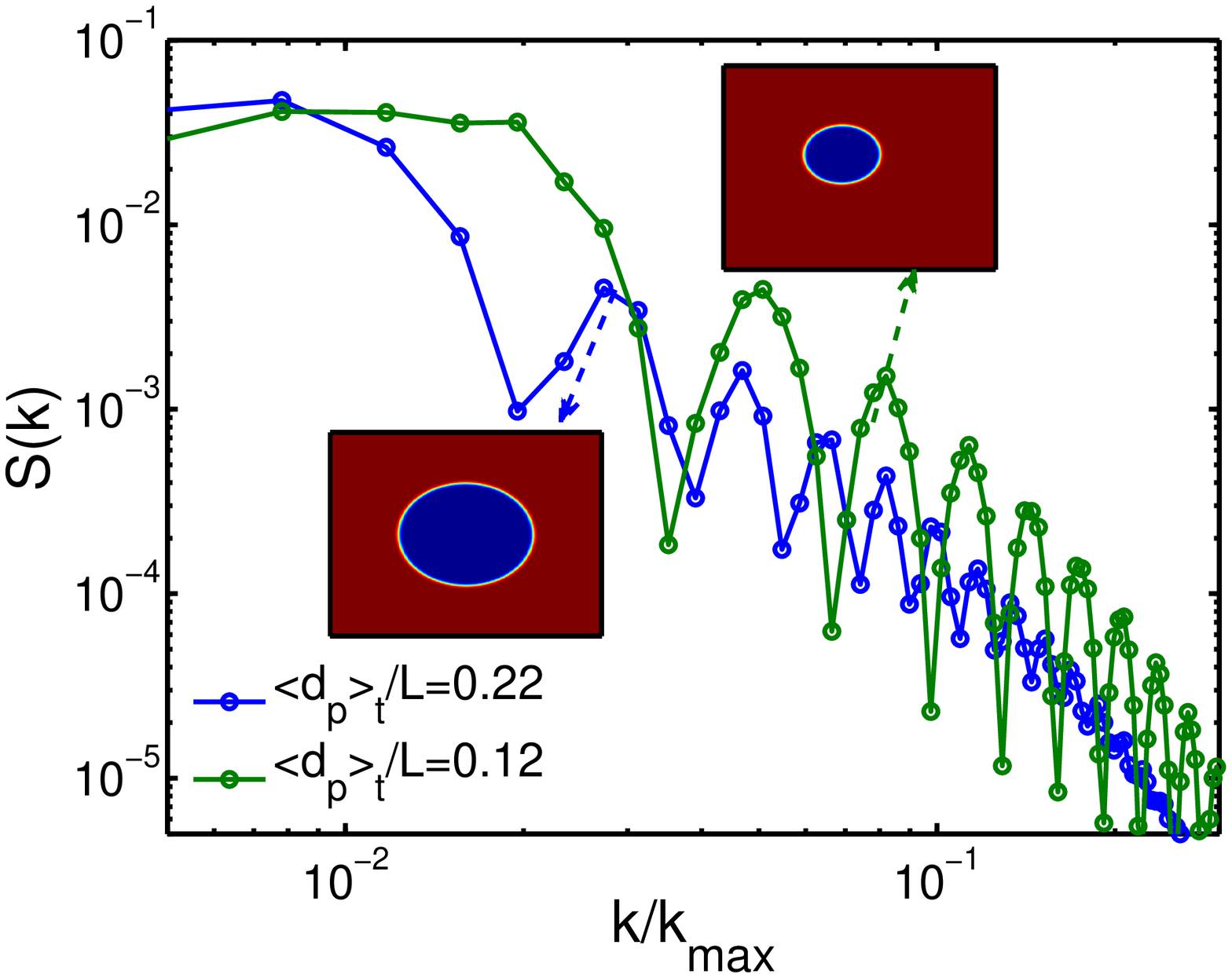}
\put(-135,110){\bf \scriptsize(f)}

\caption{(Color online) Log-log plots (base $10$) versus the scaled
wavenumber $k/k_{max}$ of (a) $E(k)$
for runs {\tt{R2}} ($\langle d_p\rangle_t/L=0.324$, deep-blue line with asterisks), 
{\tt{R12}} ($\langle d_p\rangle_t/L=0.22$, green line with crosses),
{\tt{R16}} ($\langle d_p\rangle_t/L=0.177$, red line with circles),
{\tt{R20}} ($\langle d_p\rangle_t/L=0.126$, light-blue
line with plus signs), and {\tt{R1}} (single-phase fluid, 
magenta line); the power-laws $k^{-3.6}$ and $k^{-5.2}$
are depicted by yellow-dash-dot and black-dashed lines, respectively; (b)
$E(k)$
for runs {\tt{R7}} ($\langle d_p\rangle_t/L=0.22$, $We=5.34$, green line with circles)
and a 2D Navier-Stokes run with a single-phase fluid, but with a viscosity of $\nu_{eff}(k)=\nu + \Delta \nu(k)$
(blue line with circles);
(c) plots of $\langle Re_{\lambda} \rangle_t$ 
versus $\sigma$ for
the runs {\tt{R7}}-{\tt{R14}} ($\langle d_p\rangle_t/L=0.22$) (the inset shows
$\langle Re_{\lambda} \rangle_t$ versus $\langle d_p \rangle_t/L$ for the runs 
{\tt{R2}}-{\tt{R6}}, {\tt{R12}}
 and {\tt{R16}}-{\tt{R20}} ($\sigma=0.069$ or $We=1.38$));
(d) the multifractal spectrum
$f_{\varepsilon}(\alpha)$ versus $\alpha$ of the normalized energy-dissipation
rate
$\varepsilon(t)$ versus $t$
 for $\langle d_p \rangle_t/L=0.324$ ({\tt{R2}}, blue
circles), $\langle d_p \rangle_t/L=0.15$ ({\tt{R20}}, green
squares) and single-phase fluid turbulence ({\tt{R1}}, red diamonds);
 the inset shows the corresponding normalized energy-dissipation
rate
$\varepsilon(t)$ versus $t$
for the same runs;
the order-parameter spectrum
$S(k)=|\hat{\phi}(k)|^{2}$ for the runs 
(e) {\tt{R7}} ($We=5.34$, deep-blue 
line with circles) and {\tt{R13}}
($We=0.534$, green line with circles) at $\langle d_p\rangle_t/L=0.22$; the
insets show pseudocolor plots
of $\phi$ with dotted arrows
indicating the corresponding $We$ and 
(f) {\tt{R12}} ($\langle d_p\rangle_t/L=0.22$,
deep-blue line with circles) and {\tt{R20}} ($\langle d_p\rangle_t/L=0.126$,
green line with circles); the insets show pseudocolor plots of $\phi$.}
\label{new_energy}
\end{figure*}

The inertial-range size of our droplet ensures that the 
background fluid is perturbed by it.
To explore how the droplet affects the turbulence, we first present log-log plots
of the energy spectra $E(k)$ (with and without the droplet) versus the
scaled wavenumber $k/k_{max}$, where $k_{max}=N/4$ is the maximum wavenumber in
our dealiased DNS.  We find that $E(k)$ is modified in {\it{two important
ways}} by the droplet : (1) $E(k)$ shows oscillations whose period is related
inversely to $\langle d_p\rangle_t$; (2) the large-$k$ tail of $E(k)$ is
enhanced by the droplet~\cite{spectrum_footnote}.  This enhancement is similar
to that in fluid turbulence with polymer additives~\cite{gupta15}; and it can
be understood by introducing the scale-dependent effective viscosity
$\nu_{eff}(k)=\nu+\Delta \nu(k)$ (in Fourier space), with
\begin{equation}
\Delta \nu(k) \equiv \sum\limits_{k-1/2<k^{\prime}\leq k+1/2}\frac{{\mathbf{u}}_{{\mathbf{k}}^{\prime}}.\left(\phi\bigtriangledown\mu\right)_{-{\mathbf{k}}^{\prime}}}{k^{2}E(k)}
\end{equation}
and $\left(\phi\bigtriangledown\mu\right)_{{\mathbf{k}}}$ the Fourier transform
of $\left(\phi\bigtriangledown\mu\right)$ (Eqs.~(\ref{ns})-(\ref{ch})).  In the
inset of Fig.~\ref{new_energy}(a) we plot $\Delta \nu(k)$ versus $k/k_{max}$
for the illustrative case $\langle d_p\rangle_t/L=0.324$ (deep-blue line with
asterisks); when $\Delta \nu (k) > 0$, $E(k)$ is less than its
single-phase-fluid value (magenta curve); and when $\Delta \nu(k) < 0$, $E(k)$
is greater than its single-phase-fluid value.  The change in the sign of
$\Delta \nu$ occurs at a value of $k/k_{max}$ that depends on $\langle
d_p\rangle_t/L$; the smaller the value of $\langle d_p\rangle_t/L$, the larger
is the value of $k/k_{max}$ at which $\Delta \nu(k)$ goes from being positive
to negative.  As
$\langle d_p\rangle_t/L$ increases, $E(k)$ falls less steeply with $k$ in the
power-law range; e.g., $E(k)\sim k^{-5.2}$ if there is no droplet and $E(k)
\sim k^{-3.6}$ if $\langle d_p\rangle_t/L=0.324$.  Because we use a friction
term, in the inertial range $E(k)$ scales as $\simeq k^{-5.2}$, which is
considerably different from $-3$, the exponent in the limit of no
friction~\cite{perlekar09,bofetta12}.  At low $k$, $E(k)$ decreases as $\langle
d_p\rangle_t/L$ increases.  For intermediate values of $k$, $E(k)$ decreases as
$\langle d_p\rangle_t/L$ decreases.

The large-$k$ enhancement of $E(k)$ leads to dissipation reduction, as in fluid
turbulence with polymer additives~\cite{gupta15}.  
To check
that $\nu_{eff}(k)$ can capture the effects that the droplet has on the fluid
turbulence, we have carried out some test simulations of the two-dimensional
(2D) Navier-Stokes (NS) equation, with $1024^{2}$ collocation points and the
viscosity $\nu$ replaced by  $\nu_{eff}(k)$, which we obtain from the above
equation and our DNS of the 2D CHNS equations. Clearly, our 2D NS simulation
does not have a droplet; however, it yields an energy spectrum that matches
 the one we obtain from our DNS of the 2D CHNS equations with a droplet, in a statistical sense.
We
give representative plots of energy spectra, in the steady state in
Fig.~(\ref{new_energy}(b)); these spectra agree with each other, at any given time,
for both our 2D NS and 2D CHNS runs. We conclude, therefore, that $\nu_{eff}(k)$
can capture the droplet-induced modifications of turbulent energy spectra. 
Such dissipation reduction
can be characterized by obtaining the time-series of the enstrophy or the
palinstrophy ($=\langle\frac{1}{2}\left(\bigtriangledown \times
\omega\right)^{2}\rangle$) as in Ref.~\cite{gupta15}.  Here we provide evidence
of energy-dissipation reduction as follows: when we reduce $We$ (i.e., increase
$\sigma$) with $Gr$ held fixed, the steady-state $\langle Re_{\lambda}
\rangle_t$ increases, as shown in Fig.~\ref{new_energy}(c).  $\langle Re_{\lambda}
\rangle_t$ also increases as $\langle d_p\rangle_t/L$ decreases
(Fig.~\ref{new_energy}(c) inset), because the energy required to maintain the
interface decreases as $\langle d_p\rangle_t/L$ is reduced.
In Figs.~\ref{new_energy}(d) we show, the
plot of the multifractal spectrum $f_{\varepsilon}(\alpha)$ of the energy dissipation $\varepsilon(t)/\langle \varepsilon \rangle_t$,
obtained from its time series (see inset of Fig.~\ref{new_energy}).
These plots show clearly that, because of the two-way coupling between the two fluids,
$f_{\varepsilon}(\alpha)$ is modified by the motion
of the droplet through the turbulent, background fluid.

Figure~\ref{new_energy}(a) shows oscillations in $E(k)$. Similar, but clearer,
oscillations appear in the order-parameter spectra $S(k)$, which we show in
Fig.~\ref{new_energy}(e) for $We=0.534$ and $We=5.34$ for $\langle
d_p\rangle_t/L=0.22$, and in Fig.~\ref{new_energy}(f), for $\langle d_p\rangle_t/L=0.12$
and $\langle d_p \rangle_t/L=0.22$ with $We=0.267$.  The period of these oscillations $\left(\Delta
k\right)_{osc}\simeq 2\pi/\langle d_p\rangle_t$, as we expect for such
droplets. If the fluctuations of these droplets, relative to a perfectly
circular one, are small (when $\sigma$ is large or $\langle d_p\rangle_t/L$ is
small), then the oscillations are very well defined.  We have checked that our
results do not change qualitatively if we use a higher value of $Gr$, e.g.,
$Gr=1.5\times 10^{8}$.

\section{IV. CONCLUSIONS}
Our extensive DNS of the 2D CHNS equations~(\ref{ns})-(\ref{ch}) shows that the
two-way coupling between the droplet and the background phase yields very
interesting results: The fluid turbulence leads to rich, multifractal
fluctuations in the droplet shape. Furthermore, the droplet motion modifies
$E(k)$ in two important ways : (a) oscillations with period $\simeq
2\pi/\langle d_p\rangle_t$ appear; (b) and the large-$k$ tail of $E(k)$ is
enhanced relative to that in single-fluid NS turbulence.  This enhancement can
be rationalized in terms of the scale-dependent viscosity $\nu_{eff}(k)$, which
results in dissipation reduction.  By using soap-film experiments,
Ref.~\cite{solomon98} has investigated droplet breakup in two-dimensional
chaotic flows.  Similar experiments in the turbulent regime should be able to
verify our predictions of multifractal droplet dynamics, droplet-induced
modifications of $E(k)$, and the dissipation reduction that follows from the
enhancement of the large-$k$ tail of $E(k)$.

Drag reduction by bubbles occurs in 
wall-bounded turbulent flows~\cite{Trigvason2004}; it has also been studied
in the limit of minute bubbles~\cite{proc2005}.
We show that, even at the level of a single droplet with a diameter
in inertial-range scales, we obtain the bulk analog
of drag reduction, namely, dissipation reduction in homogeneous, isotropic
turbulence. Furthermore, the analog of the large-$k$ enhancement in $E(k)$,
which we find here, has been seen in three-dimensional experiments
in turbulent bubbly flows~\cite{lohse06,prakash13,mendez13}.

Although the CHNS approach has been used to study droplet dynamics in a
laminar~\cite{prosperetti2014,bifarale2014,prakash2015} flow, wall-drag of a
droplet in a turbulent channel flow~\cite{scarbolo2015a}, droplet breakup or
coalescence~\cite{scarbolo2015}, steady-state droplet-size
distributions~\cite{perlekar12,skartlien13}, and the turbulence-induced arrest
of phase separation~\cite{perlekar2014}, it has neither been used to study
droplet fluctuations and droplet-acceleration statistics, in a turbulent flow,
nor the modification of fluid turbulence by droplet fluctuations because of the
two-way coupling, which we investigate. These issues have also not been
considered by other DNSs of drag reduction in channel flows~\cite{xu2002},
boundary layers~\cite{ferrante99,jacob2010}, and in some
experiments~\cite{hassan2002,fontainne99} with droplets. 

\begin{acknowledgements}

We thank S.S.Ray for discussions.  NP and RP thank SERC (IISc) for
computational resources, the Department of Science and Technology and the
University Grants Commission (India) for support; PP thanks the Department of
Atomic Energy (India); AG thanks a grant from the European Research Council
(ERC) under the European Community's Seventh Framework Programme
(FP7/2007-2013)/ERC Grant Agreement No.279004.

\end{acknowledgements}

\section{APPENDIX}
\begin{figure*}
\includegraphics[width=.45\linewidth]{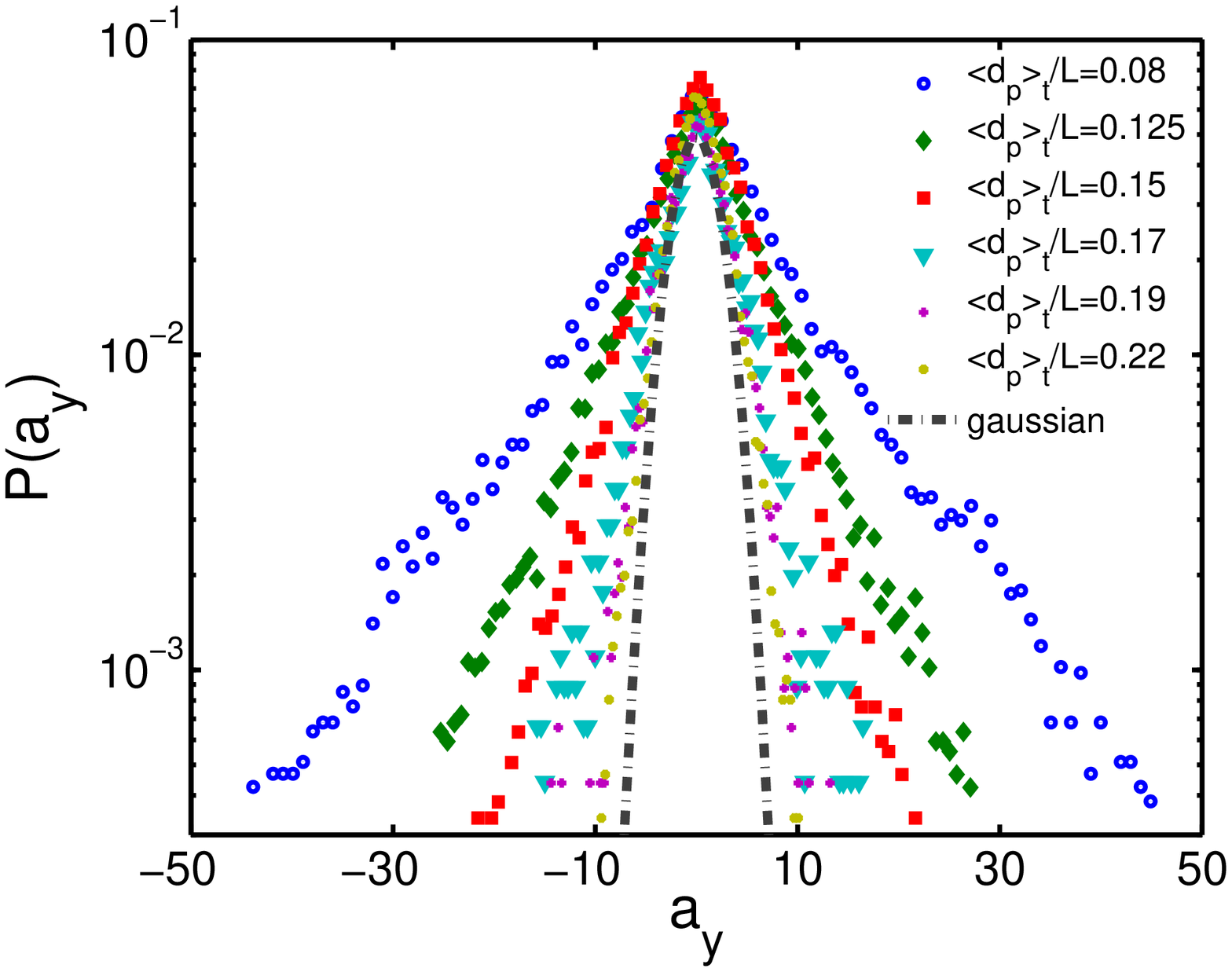}\
\put(-185,160){\bf (a)}
\includegraphics[width=.45\linewidth]{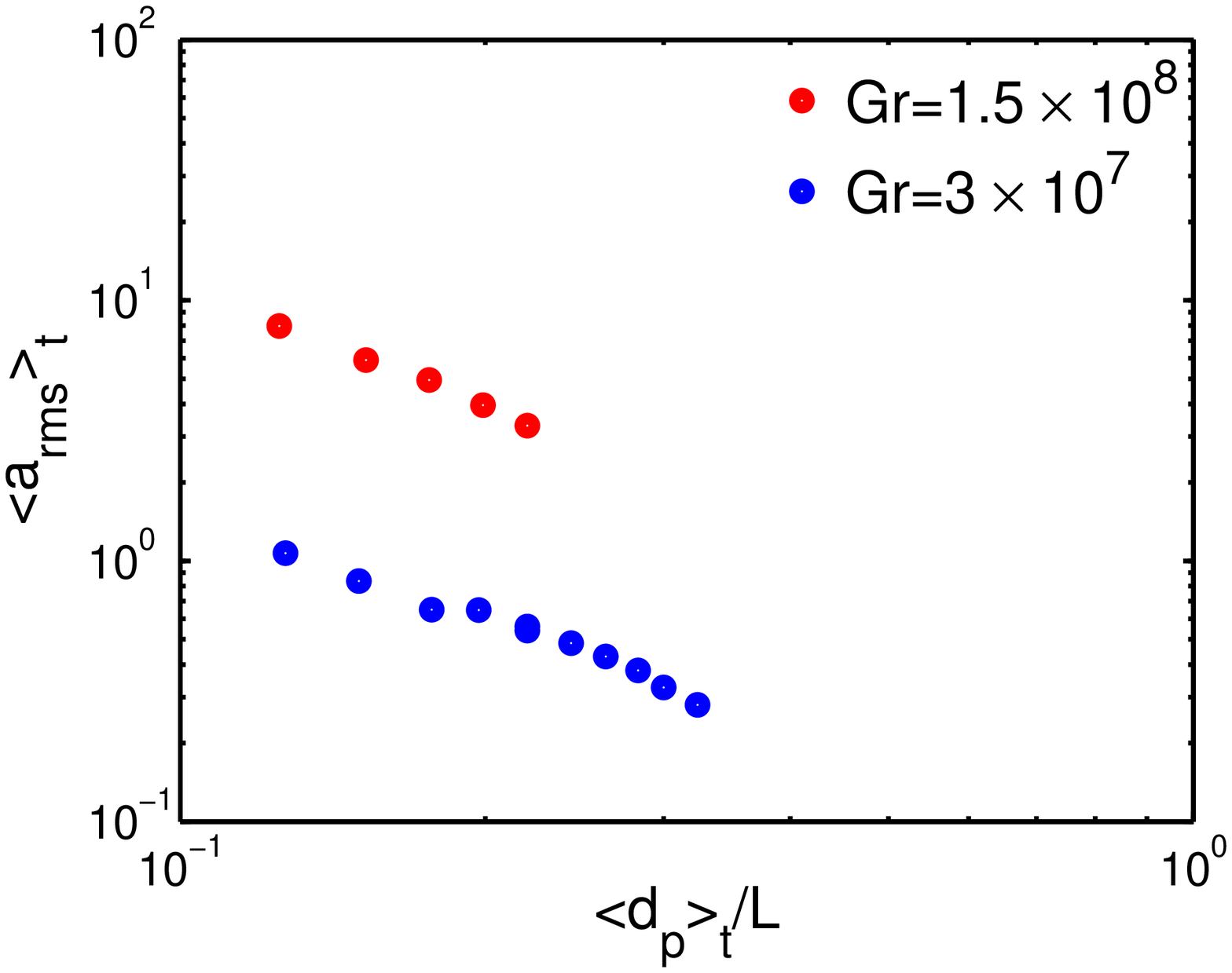}\
\put(-185,160){\bf (b)}
\caption{(Color online)(a) Semilog (base $10$) plots of the PDFs  $P(a_{y})$, the PDF of $a_{y}$ of the center of mass of the 
 droplets for runs {\tt{R28}} ($\langle d_p\rangle_t/L=0.126$, deep-blue circles), {\tt{R27}} ($\langle d_p\rangle_t/L=0.153$, green diamonds),
{\tt{R26}} ($\langle d_p\rangle_t/L=0.22$, red squares), {\tt{R25}} ($\langle d_p\rangle_t/L=0.263$, light-blue downward-pointing triangles), 
{\tt{R24}} ($\langle d_p\rangle_t/L=0.283$, magenta plus signs) and {\tt{R23}} ($\langle d_p\rangle_t/L=0.32$, yellow asterix)
 at $We=0.138$ (these
PDFs are not scaled by their rms values); (b) log-log (base $10$) plot $\langle a_{rms}\rangle_t$, the  root-mean square acceleration of 
the droplet center of mass, versus $\langle d_p\rangle_t/L$. }
\label{highGr}
\end{figure*}

In the main part of this paper we 
have presented results for $Gr=3\times10^{7}$.
We show now that these results are
qualitatively unchanged when we increase $Gr$ to, say, $Gr=1.5 \times 10^{8}$.
Consider, e.g., the illustrative plot of $P(a_y)$ versus
$a_y$ for $Gr=1.5 \times 10^{8}$ that we show in Fig.~\ref{highGr}(a).
This is qualitatively similar to Fig.~\ref{spectra}(b) for $Gr=3\times 10^{7}$.
In Fig.~\ref{highGr}(b) we show the plots of $\langle a_{rms} \rangle_t$ versus $\langle d_p \rangle_t/L$
for $Gr=3 \times 10^{7}$ and $Gr=1.5 \times 10^{8}$; although
the curve for $Gr=1.5 \times 10^{8}$ lies well above that for $Gr=3\times 10^{7}$.

In the multifractal spectrum calculation, we use a Wavelet Transform Modulus
Maxima Method. 
The wavelet transform of a function $f$ decomposes it into
several elementary wavelets, which are all constructed from a single
the analysing wavelet $\psi$. This transform is defined as follows:
\begin{equation} T_{\psi}[f](x,a) = \frac{1}{a}\int
\limits_{-\infty}^{+\infty}\psi (\frac{x-b}{a})f(x)dx ,
\end{equation} 
where $a\in \mathcal{R}$ is a scale parameter and $b\in\mathcal{R}$ is a space
parameter; structures smaller than $a$ are smoothed out; and the wavelet $\psi$
is invariant under spatial shifts of length $b$.
At each scale $a$, we pick the local maxima of $|T_{\psi}f(x,a)|$ and
define the following partition function:
\begin{equation}
Z(a,q)=\sum \limits_{l\in \mathcal{L}(a)} \left(\sup_{(x,a^{\prime})\in l} |T_{\psi}f(x,a^{\prime})|\right)^{q},
\end{equation}
where $q\in \mathcal{R}$.  In the limit $a\rightarrow 0$, the Renyi
exponents $\tau(q)$ follow from
\begin{equation}
Z(a,q)\sim a^{\tau(q)};
\end{equation}
the following Legendre transform of $\tau(q)$ yields the multifractal spectrum
\begin{equation}
f(\alpha)=\min _{\alpha}[q\alpha - \tau(q)],
\end{equation}
where $\alpha=d\tau(q)/dq$. 
 In our calculations we follow
Ref.~\cite{muzy1993}; in particular, we use a slightly modified version of the
computer program given in Refs.~\cite{ashken,physionet}.
In our calculations, the analyzing wavelet is a Gaussian function.
We obtain partition functions $Z(a,q)$ between moments $q_{max}$ and $q_{min}$, with resolution
$dq$, $q_{max}=2.0$, $q_{min}=-2.0$, and $dq=0.2$. The value of $a$ is $L_s/8$,
where $L_s$ is the signal length.


\begin{thebibliography}{10}




\bibitem{jeremie2006}
J. Bec, L. Biferale, G. Boffeta , A. Celani,
M. Cencini , A. Lanotte , S. Musacchio
and F. Toschi, J. Fluid Mech., {\bf 550}, 349-358 (2006).

\bibitem{grabowski2013}
W. W. Grabowski and L.P. Wang, Annu. Rev. Fluid Mech. {\bf 45}, 293-324, (2013).
\bibitem{shaw2003}
R. A. Shaw, Annu. Rev. Fluid Mech. {\bf 35.1}, 183-227 (2003).
\bibitem{fuel}
C.A. Chryssakis, D. Assanis, J. Lee and K. Nishida, 
No. 2003-01-0007, SAE Technical Paper, 2003.
\bibitem{baroud2010}
C. N. Baroud, F. Gallaire, and R. Dangla, Lab on a Chip {\bf 10}, 2032-2045 (2010).
\bibitem{inkjet}
A van der Bos, M J van der Meulen, T. Driessen, M van den Berg, H. Reinten, H. Wijshoff, M. Versluis, and D. Lohse,
Phys. Rev. Applied {\bf 1}, 014004 (2014).
\bibitem{lohse2003}
I.M. Mazzitelli, D. Lohse and F. Toschi, Phys. Fluids {\bf 15}, L5-L8 (2003). 
\bibitem{bifarale2005}
L. Biferale, G. Boffeta, A. Celani, A. Lanotte,
and F. Toschi, Phys. Fluids {\bf 17} 021701 (2005).


\bibitem{chaikan2000}
P. M. Chaikin and T. C. Lubensky, {\em Principles of Condensed Matter Physics}  (Cambridge University Press,
 Reprint edition (2000)) .

\bibitem{halperin77} P.C. Hohenberg and B. I. Halperin, Rev.Mod. Phys {\bf 49} 435 (1977).
\bibitem{lifshitz59} I. M. Lifshitz and V. V. Slyozov, J. Phys. Chem. Solids
{\bf 19}, 35 (1959); H. Furukawa, Phys. Rev. A {\bf 31}, 1103
(1985); E. D. Siggia, Phys. Rev. A {\bf 20}, 595 (1979);

\bibitem{domb83}
J.D. Gunton, M. San Miguel, and P.S. Sahni, in
{\em Phase Transitions and Critical Phenomena}, eds.
C. Domb and J.L. Lebowitz, Vol. {\bf 8} (Academic, London, 1983).
\bibitem{bray94} A. J. Bray,
Adv. Phys., {\bf 43}, 357-459, 1994.

\bibitem{nucleation} J. Lothe and G.M. Pound, J. Chem. Phys. {\bf 36}, 2080
(1962).


\bibitem{onuki2002} A. Onuki, {\em Phase Transition Dynamics} (Cambridge University
Press, UK, 2002).

\bibitem{badalassi2003}
V. E. Badalassi, H. D. Ceniceros, and S. Banerjee, J. Comput. Phys. {\bf 190}, 371–397 (2003).

\bibitem{perlekar2014}
P. Perlekar, R. Benzi, H. J. H. Clercx, D. R. Nelson and F. Toschi, Phys. Rev. Lett, {\bf{112}}, 014502 (2014). 

\bibitem{cahn61}
J.W. Cahn, Acta metall {\bf 9}, 795 1961.

\bibitem{bofetta2005}
S. Berti, G. Boffetta, M. Cencini and A. Vulpiani, Phys. Rev. Lett. {\bf 95}, 224501 (2005).
A.J. Wagner and J. M. Yeomans, Phys. Rev. Lett. {\bf 80}, 1429
(1998); V.M. Kendon, Phys. Rev. E, {\bf 61}, R6071 (R) (2000);
V.M. Kendon, M.E. Cates, I.P. Barraga, J.C. Desplat, P. Blandon, J. Fluid Mech., {\bf 440}, 147 (2001);
S. Puri, in
{\em Kinetics of Phase Transitions}, eds. S. Puri and V.
Wadhawan (CRC Press, Boca Raton, US, 2009), Vol. {\bf 6}, p. 437.

\bibitem{bofetta2009} G. Boffetta and R. Ecke, Annu. Rev Fluid Mech. {\bf 44},
427-451 (2012); R. Pandit, P. Perlekar, and S. S. Ray,
Pramana-Journal of Physics, {\bf 73}, 157(2009).

\bibitem{prasad2011}
P. Perlekar, S. S. Ray, D. Mitra and R. Pandit, Phys.
Rev. Lett. {\bf 106}, 054501 (2011).

\bibitem{fjortoff}
R. Fj{\o}rtoft, Tellus {\bf 5}, 226 (1953).
\bibitem{Kraich}
R. H. Kraichnan, Phys. Fluids {\bf 10}, 1417 (1967).
\bibitem{Leith}
C. Leith, Physics of Fluids {\bf 11}, 671 (1968).
\bibitem{Batchelor}
G. K. Batchelor, Phys. Fluids Suppl. II {\bf 12}, 233 (1969).
\bibitem{leisure}
M. Lesieur, {\em Turbulence in Fluids}, Vol. 84 of 
{\em Fluid Mechanics and Its Applications} (Springer, The Netherlands,
2008)
\bibitem{jeremie}
H. Homann and J. Bec, J. Fluid Mech., {\bf 651}, 81-91 (2010).
\bibitem{qureshi2007}
N.M. Qureshi, M. Bourgoin, C. Baudet, A. Cartellier, Y. Gagne,
Phys. Rev. Lett. {\bf 99}, 184502 (2007).
\bibitem{hill}
R. J. Hill and J.M. Wilczak, J. Fluid Mech., {\bf 296}, 247–269 (1995).
\bibitem{kalelkar05}
C. Kalelkar, R. Govindarajan, and R. Pandit, Phys. Rev.
E {\bf 72}, 017301 (2005).

\bibitem{perlekar06}
P. Perlekar, D. Mitra, and R. Pandit, Phys. Rev. Lett. {\bf 97}, 
264501 (2006); W.H.Cai, F.C.Li and H.N. Zhang, J. Fluid Mech. {\bf 665}
334 (2010).


\bibitem{gupta15}
A. Gupta, P. Perlekar, R. Pandit, Phys. Rev. E {\bf 91}(3), 033013
(2015).
 



\bibitem{celani2009}
A. Celani, A. Mazzino, P. Muratore--Ginanneschi and L. Vozella, J.Fluid Mech., {\bf{622}}, 115-134 (2009). 
\bibitem{scarbolo2013}
L. Scarbolo and A. Soldati, J. Turb. {\bf{14}}, 11 (2013). 

\bibitem{shen2004}
P. Yue, J.J. Feng, C. Liu, and J. Shen, J. Fluid Mech. {\bf 515}, 293-317 (2004).

\bibitem{scarbolo2011}
L. Scarbolo, D. Molin and A. Soldati,
 APS Division of Fluid Dynamics Meeting Abstracts.  {\bf 1}, 4002 (2011).
\bibitem{kolmogorov}
S. Childress, R.R. Kerswell and A.D. Gilbert, Physica D {\bf 158}, 105-128 (2001).
\bibitem{spectral}
 C. Canuto, M. Y. Hussaini, A. Quarteroni,  and T. A. Zang,  {\em Spectral Methods in Fluid
Dynamics}, Springer.
\bibitem{cox2002}
S. M. Cox, and P. C. Matthews, J. Comput. Phys. {\bf 176}, 430-455 (2002).
\bibitem{cuda}
{\em http://www.nvidia.com/object/cuda$\_$home$\_$new.html}.
\bibitem{perlekar12}
P. Perlekar, L. Biferale, M. Sbragaglia, S. Srivastava, and
F. Toschi, Phys. Fluids {\bf 24}, 065101 (2012).

\bibitem{suppmat}
\burl {https://www.youtube.com/watch?v=p-lXR9VRcjI&feature=youtu.be}. \burl {https://www.youtube.com/watch?v=DxspQUL46pU&feature=youtu.be}.

\bibitem{muzy1993}
J. F. Muzy, E. Bacry, and A. Arneodo, Phys. Rev. E {\bf 47},
875 (1993).
\bibitem{ashken}
https://www.physionet.org/physiotools/multifractal
\bibitem{physionet}
A.L. Goldberger, L.A.N Amaral, L. Glass, J.M. Hausdorff, Pch Ivanov, R.G. Mark, J.E. Mietus, G.B. Moody, 
C-K Peng, H.E. Stanley, PhysioBank, PhysioToolkit, and PhysioNet,
Components of a New Research Resource for Complex Physiologic Signals. Circulation {\bf 101}:e215-e220
 [Circulation Electronic Pages; http://circ.ahajournals.org/cgi/content/full/101/23/e215]; 2000.
\bibitem{biferale13conf}
L. Biferale, P. Perlekar, M. Sbragaglia, S. Srivastava, and F. Toschi,
J. Phys.: Conf. Ser. {\bf 318}, 052017 (2011).

\bibitem{jeremie_footnote}
Even in the 3D studies
of Refs.\cite{jeremie,qureshi2007}, the power-law ranges are small.
\bibitem{spectrum_footnote}
In the absence of this droplet, our
forcing scheme yields a fluid-energy spectrum that is dominated by a forward
cascade of the enstrophy.
\bibitem{perlekar09}P. Perlekar and R. Pandit, New
 J.
 Phys., {\bf 
11}, 073003
(2009).
\bibitem{bofetta12}
G. Boffetta and R. E. Ecke, Annu. Rev. Fluid Mech. {\bf 44},
427 (2012).
\bibitem{solomon98}
T. H. Solomon, S. Tomas, and J. L. Warner, Phys. Fluids
{\bf 10}, 342 (1998).




\bibitem{Trigvason2004}
J. Lu, and Gretar Tryggvason, APS Division of Fluid Dynamics Meeting Abstracts, {\bf 1} (2004).

\bibitem{proc2005}
V. S. L\`vov, A. Pomyalov, I. Procaccia, and V. Tiberkevich, 
Phys. Rev. Lett. {\bf 94}, 174502 (2005).

\bibitem{lohse06}
T. H. Van Den Berg, S. Luther, I. M. Mazzitelli,
J. M. Rensen, F. Toschi and D. Lohse,
Journal of Turbulence {\bf 7}, No. 14, 2006.




\bibitem{prakash13}
V.N. Prakash, J.M. Mercado,
F.E.M. Ramos, Y. Tagawa,
D. Lohse and C. Sun, arXiv preprint arXiv:1307.6252 (2013).

\bibitem{mendez13}
S. Mendez-Diaz, J. C. Serrano-Garcia, R. Zenit,
and J. A. Hernandez-Cordero, Phys. Fluids, {\bf 25}, 043303 (2013).

\bibitem{prosperetti2014}
B. Ray and A. Prosperetti, Chemical engineering science {\bf 108}, 213-222 (2014).
\bibitem{bifarale2014}
L. Biferale, C. Meneveau  and R. Verzicco, J. Fluid Mech. {\bf 754}, 184–207 (2014).
\bibitem{prakash2015}
N.J. Cira, A. Benusiglio, and M. Prakash, Nature {\bf 519}, 446–450 (2015).
\bibitem{scarbolo2015a}
L. Scarbolo, A. Soldati, Comput. $\&$ Fluids {\bf 113}, 87-92 (2015).
\bibitem{scarbolo2015}
L. Scarbolo, F. Bianco, A. Soldati, Phys. Fluids {\bf 27}, 073302 (2015).
\bibitem{skartlien13}
R. Skartlien, E. Sollum, H. Schumann,
J. Chem. Phys. {\bf 139}, 174901 (2013).

\bibitem{xu2002}
J. Xu, M. Maxey and G. Karniadakis, J. Fluid Mech. {\bf 468}, 271–281 (2002).
\bibitem{ferrante99}
A. Ferrante and S. Elghobashi, J. Fluid Mech. {\bf 503}, 345 (1999).

\bibitem{jacob2010}
B. Jacob, A. Olivieri, M. Miozzi, E. F. Campana, and R. Piva, Phys. Fluids {\bf 22}, 115104 (2010).
\bibitem{hassan2002}
Y.A. Hassan and J. Ortiz-Villafuerte, 
In Proceedings 11th Int. Symp. Applications of Laser Techniques to
Fluid Mechanics Lisbon, July 8-11, (Paper 23.3), (2002).
\bibitem{fontainne99}
A.A. Fontaine, S. Deutsch, T.A. Brungart, H.L. Petrie and
M. Fenstermacker, Exp. Fluids, {\bf 26} 397–403 (1999).























\end{thebibliography}
\end{document}